\documentclass[]{emulateapj}

\usepackage{amsmath,mathtools,float,subfigure}
\usepackage{color,ulem,epstopdf}
\usepackage{graphicx}

\newcommand{\w}{}

\newcommand{\bc}{}
\newcommand{\bcc}{}

\def\refnew#1{(\ref{#1})}

\shorttitle{The Evaporation Valley}
\shortauthors{Owen, J. E. \& Wu, Y.}

\begin{document}

%% LaTeX will automatically break titles if they run longer than
%% one line. However, you may use \\ to force a line break if
%% you desire.

\title{The evaporation valley in the {\it Kepler} planets}
% the lens to day one
%The origin of the evaporation gap}

%% Use \author, \affil, and the \and command to format
%% author and affiliation information.
%% Note that \email has replaced the old \authoremail command
%% from AASTeX v4.0. You can use \email to mark an email address
%% anywhere in the paper, not just in the front matter.
%% As in the title, use \\ to force line breaks.

\author{James E. Owen\altaffilmark{1}}
\affil{Institute for Advanced Study, Einstein Drive, Princeton, NJ 08540, USA}
\email{jowen@ias.edu}
%\and
\author{Yanqin Wu}%\altaffilmark{2}}
\affil{Department of Astronomy and Astrophysics, University of Toronto, Toronto, ON M5S 3H4, Canada}
\email{wu@astro.utoronto.ca}
%
%\and
%
%\author{R. J. Hanisch\altaffilmark{5}}
%\affil{Space Telescope Science Institute, Baltimore, MD 21218}

%% Notice that each of these authors has alternate affiliations, which
%% are identified by the \altaffilmark after each name.  Specify alternate
%% affiliation information with \altaffiltext, with one command per each
%% affiliation.

\altaffiltext{1}{Hubble Fellow}
%\altaffiltext{2}{Society of Fellows, Harvard University.}
%\altaffiltext{3}{present address: Center for Astrophysics,
%    60 Garden Street, Cambridge, MA 02138}
%\altaffiltext{4}{Visiting Programmer, Space Telescope Science Institute}
%\altaffiltext{5}{Patron, Alonso's Bar and Grill}

%% Mark off your abstract in the ``abstract'' environment. In the manuscript
%% style, abstract will output a Received/Accepted line after the
%% title and affiliation information. No date will appear since the author
%% does not have this information. The dates will be filled in by the
%% editorial office after submission.

\begin{abstract} 
A new piece of evidence supporting the photoevaporation-driven evolution model for low-mass, close-in exoplanets was recently presented by the California-Kepler-Survey.  The radius distribution of the {\it Kepler} planets is shown to be bimodal, with a ``valley' separating two peaks at $1.3$ and $2.6\,$R$_\oplus$. Such an ``evaporation-valley' had been predicted by numerical models previously. Here, we develop a minimal model to demonstrate that this valley results from the following fact: the timescale for envelope erosion is the longest for those planets with hydrogen/helium-rich envelopes that, while only a few percent in weight, double its radius. The timescale falls for envelopes lighter than this because the planet's radius remains largely constant for tenuous envelopes. The timescale also drops for heavier envelopes because the planet swells up faster than the addition of envelope mass.  Photoevaporation therefore herds planets into either bare cores ($\sim1.3\,{\rm~R}_\oplus$), or those with double the core's radius ($\sim2.6\,{\rm~R}_\oplus$). This process mostly occurs during the first $100\,$Myrs when the stars' high energy flux are high and nearly constant. The observed radius distribution further requires that the {\it Kepler} planets are clustered around $3\,{\rm~M}_\oplus$~in mass, are born with H/He envelopes more than a few percent in mass, and that their cores are similar to the Earth in composition. Such envelopes must have been accreted before the dispersal of the gas disks, while the core composition indicates formation inside the ice-line. Lastly, the photoevaporation model fails to account for bare planets beyond $\sim30-60\,$days, if these planets are abundant, they may point to a significant second channel for planet formation, resembling the Solar-System terrestrial planets.
  \end{abstract}

%% Keywords should appear after the \end{abstract} command. The uncommented
%% example has been keyed in ApJ style. See the instructions to authors
%% for the journal to which you are submitting your paper to determine
%% what keyword punctuation is appropriate.

\keywords{planets and satellites: atmospheres --- planets and satellites: composition --- planets and satellites: formation --- planets and satellites: physical evolution}

%% From the front matter, we move on to the body of the paper.
%% In the first two sections, notice the use of the natbib \citep
%% and \citet commands to identify citations.  The citations are
%% tied to the reference list via symbolic KEYs. The KEY corresponds
%% to the KEY in the \bibitem in the reference list below. We have
%% chosen the first three characters of the first author's name plus
%% the last two numeral of the year of publication as our KEY for
%% each reference.

%% Authors who wish to have the most important objects in their paper
%% linked in the electronic edition to a data center may do so by tagging
%% their objects with \objectname{} or \object{}.  Each macro takes the
%% object name as its required argument. The optional, square-bracket 
%% argument should be used in cases where the data center identification
%% differs from what is to be printed in the paper.  The text appearing 
%% in curly braces is what will appear in print in the published paper. 
%% If the object name is recognized by the data centers, it will be linked
%% in the electronic edition to the object data available at the data centers  
%%
%% Note that for sources with brackets in their names, e.g. [WEG2004] 14h-090,
%% the brackets must be escaped with backslashes when used in the first
%% square-bracket argument, for instance, \object[\[WEG2004\] 14h-090]{90}).
%%  Otherwise, LaTeX will issue an error. 

\section{Introduction}

Recent exoplanet discovery missions and targeted follow-up campaigns have fundamentally changed our understanding of what constitutes a ``typical'' planet \citep[e.g.][]{Borucki2011,Marcy2014}. Specifically, the most common type of exoplanets are smaller than Neptune ($\lesssim$4~R$_\oplus$, e.g., \citealt{Youdin2011,Howard2012,Batalha2013,Petigura2013,Burke2014,Morton2016}) and have masses of a few to tens of Earth masses. The fraction of sun-like stars that host {\bc at} least one of these ``{\it Kepler}'' planets with an orbital period of less than 100~days is around 60--90\% \citep[e.g.][]{Fressin2013,Silburt2015,Mulders2016}.

Combining transit measurements of a planet's radius with a measurement of its mass from transit-timing variations (TTVs, \citealt{Carter2012,Wu2013,Hadden2014,JontofHutter2016,Hadden2017}) or radial-velocity (RV) follow-up \citep[e.g.][]{Marcy2014,Weiss2014} quickly told us that many of these planets had compositions unlike the small terrestrial planets in our own solar system \citep[e.g.][]{Wolfgang2016}. Rather than being completely solid, such planets are likely to be composed of a dense solid core surrounded by a voluminous volatile rich envelope. However, due to degeneracies present in the mass-radius plane at low-masses \citep[e.g.][]{Adams2008,Rogers2010}, we cannot infer the compositions just based on the current measured mass and radius for the majority of the observed planets. These degeneracies can be broken by considering how the local environment effects the  {\it evolution} of the planet. For example, \citet{Wu2013} showed that closer-in planets tend to be denser, while \citet{Ciardi2013} demonstrated that for pairs of planets in multi-planet systems, the inner planet tends to be smaller. These studies suggest that the envelopes are rich in hydrogen/helium.
%We do know that some planets are so low density ($\rho\sim 0.1\,$g~cm$^{-3}$) that they must contain large H/He envelopes \citep[e.g.][]{Wu2013,JontofHutter2016,Hadden2017}, while others are so high density ($\rho \sim 6\,$g~cm$^{-3}$) that they have no volatile envelope and are consistent with an Earth-like composition \citep{Dressing2015,Lopez-Morales2016}. 
%However, it still remains unclear whether H/He envelopes constitutes the majority, or the minority with volatile-rich, silicate-rich or iron-rich cores surrounded by steam or H/He envelopes still being consistent with a large fraction of the observed planet population.

Orbiting close to their parent stars, the {\it Kepler} planets can receive, over a lifetime, an integrated high-energy irradiation (high-energy exposure) that is an appreciable fraction of their gravitational binding energy \citep[e.g.][]{Lammer2003,Lecavelier2007,Davis2009}, {\bc where this high-energy ``exposure'' is dominated by the first $\sim 100\,$Myr of the planet's lifetime \cite[e.g.][]{Jackson2012}}. Planets with H/He-rich envelopes can be strongly evaporated by this irradiation \citep[e.g.][]{Yelle2004,Tian2005,MurrayClay2009,Owen2012,Johnstone2015,Erkaev2016}, and indeed H/He evaporation has been observed from the low-mass planet GJ~436b \citep{Kulow2014,Ehrenreich2015}. Evaporation naturally results in planets that are smaller and denser than those at large separations \citep[e.g.][]{Lopez2012,Owen2013,Lopez2013,Jin2014,Howe2015}. For a planet with a low enough mass and a close enough orbit, its initial low-mass H/He envelope can even be entirely stripped, leaving behind a naked solid core. This explains the dearth of planets with any envelopes at short periods \citep{Lundkvist2016}.
The core's mass and density play a primary role in controlling a planet's evolution by setting the escape velocity \citep{Owen2013,Lopez2013,Owen2016,Zahnle2017}, 
allowing one to break the compositional degeneracies in individual systems by statistical modelling 
%of a planet's initial composition 
\citep{Owen2016}. 
%Evaporation is important for low-mass planets, whose envelopes contain less mass than any solid core %\citep[e.g.][]{Baraffe2006,Owen2012} and is thought to drive the evolution of low-mass H/He-rich %planets \citep{Owen2013,Lopez2013}. and 
While steam atmospheres are also expected to lose mass (as in the case of early Venus -- \citealt{Kasting1983}) the evaporative histories of such planets are significantly different to those that contain H/He envelopes  and therefore distinguishable \citep{Lopez2016}.

{While the evaporation theory naturally explains why {\it Kepler} planets are larger and less dense further out, it also makes another major prediction: the existence of }
an ``evaporation valley'',  a low-residence region in the radius-period plane between planets that have been completely stripped and those that are able to retain an envelope with roughly $\sim 1$\% in mass. The evaporation valley was first predicted by \citet{Owen2013} using numerical evolutionary studies for low-mass planets with pure rock (silicate) cores, and shortly after, by \citet{Lopez2013} for different core compositions using a different evaporation model. The evaporation valley was further reproduced by \citet{Jin2014} and \citet{Chen2016}, again using different evaporation prescriptions and initial populations. The evaporation valley is thus a robust prediction of evaporative driven evolution of close-in H/He rich planets. This feature is largely independent of the assumed H/He evaporation model (energy-limited, recombination limited, UV driven, X-ray driven etc.), and its creation also appears insensitive to the choice of initial conditions (for at least a reasonable range of starting conditions); however, the details of these choices do control it's properties (width, location with orbital period etc.).

The predicted occurrence valley, between stripped cores and those that retain a residual H/He envelope, is not particularly wide, with a radius width  {\bc of $\sim 0.5\,$R$_\oplus$} (see Fig.~8 of \citealt{Owen2013} and Fig.~9 of \citealt{Lopez2013}). Large planetary radius errors, stemming from uncertainties in the stellar radius, have previously hampered efforts to observationally solidify its presence in the observed exoplanet population {\citep[see][for a preliminary analysis that suggested its presence]{Owen2013}}. Recently, the California-Kepler Survey (CKS), using spectroscopic follow-up of a large (1305) sample of planet hosting {\it Kepler} stars \citep{Petigura2017} refined the planet parameters for 2025 {\it Kepler} planets, and reduced the typical planetary radius error to $\sim 10$\% \citep{Johnson2017}. The CKS sample allowed \citet{Fulton2017} to definitely reveal 
%both a gap in the radius distribution of close-in (Periods $< 100$~days) and a 
a valley in the planet occurrence rate in the planet radius -- period plane: close-in planets predominantly have a radius of either $\sim 1.3\,$R$_{\oplus}$ or $2.6\,$R$_\oplus$, while planets with a radius of $1.8\,$R$_\oplus$ are considerably rarer, in spectacular agreement with the predicted evaporation driven evolution scenario for close-in exoplanets \citep{Owen2013,Lopez2013,Jin2014,Chen2016}.        

While the presence of the evaporation valley in numerical models is robust to changes in model assumptions, a clear physical, first-principle  description of its origin is missing, along with an understanding of how its properties change with model assumptions such as the core composition and evaporation model, and even when or whether it could be made to disappear. This paper serves two purposes: firstly, we clearly explain the physics behind the origin of the evaporation-valley; secondly, we will perform a preliminary investigation into how the {\it observed} evaporation valley can break many of the composition degeneracies and make inferences about both the composition of the {\it Kepler} planets and how/where they formed.   

% Out of date now: Given the median planetary radius uncertainty in the {\it Kepler} planets is $\sim 30\%$\footnote{Primarily from uncertainties in the host stars radius} at the time of writing (using the \citet{Morton2016} sample), the evidence for and or against the presence of the evaporation valley in the observed exoplanet sample is limited as it will be washed out by the large radius errors, although \citet{Owen2013} did present weak $\sim 1.5-2\sigma$ for a bimodality in the observed planet radius distribution with planet radii of $\sim 1.7\,$R$_\oplus$ and $\sim 2.4\,$R$_\oplus$ being preferable. The smaller planets being associated with higher X-ray exposures and the larger with lower X-ray exposures.

%While the presence of the evaporation valley in numerical models appears to be robust to changes in model assumptions, a clear {\it physical} description of its origin is missing, along with an understanding of how its properties change with model assumptions such as the core composition and evaporation model, and even when or whether it could be made to disappear. In this work, we eschew numerical simulations of planetary evolution and provide a first principles description of the origin of the evaporation valley, derive its dependence on model parameters analytically and discuss its robustness to a wide range of conditions. We will start by building low-mass planetary structure models in Section~2, blah blah blah.    

\section{A {\bc minimal} analytical model}

Here, we build a minimal model of an evolving planet under the influence of evaporation and cooling. The planet is assumed to consist of a solid core 
%(of arbitrary composition but fixed density, $\rho_{c}$), with a 
of mass $M_c$, and radius $R_c$, surrounded by a gaseous envelope whose equation of state can be described by the ideal gas law. Such a model allows us to determine the planet outer radius as a function of time and envelope mass.
We then perform numerical checks to confirm these results. While a fully numerical study is straight-forward (and has been done), such an approach obscurates the underlying properties of low-mass exoplanet atmospheres that results in the evaporation valley. Furthermore, the analytical method sheds light on many of the parameter dependencies.

\subsection{Radius versus envelope mass for a low-mass planet}

Our goal here is to understand how the radius of the planet ($R_p$) changes with its envelope mass fraction ($X = M_{\rm env}/M_c$) with a view to calculating the mass-loss timescale. {\bcc  We consider low-mass envelopes ($X < 1$), so the planet's mass is still dominated by the core. Therefore, we  neglect the self-gravity of the planet's envelope.}\footnote{\bcc The role of self-gravity is to compress the envelope giving it a smaller radius than our following analysis predicts. Self-gravity becomes important when $X\sim 1$ and larger, and its impact is demonstrated in Fig~1.}

Due to stellar insolation, the envelope is adiabatic (convective) in the deep interior and roughly isothermal (and radiative) near the surface. The roughly isothermal radiative cap satisfies $T \sim T_{\rm eq}$ \citep[e.g.][]{Rafikov2006,Lee2015,Ginzburg2016},
 where the {\bc photospheric} equilibrium temperature is set by stellar insolation. The radiative-convective boundary is assumed to occur at a density $\rho_{\rm rcb}$, and a radius $R_{\rm rcb}$. The planet's radius
is set by the photospheric radius and is typically $\sim 6$ pressure scale heights above $R_{\rm rcb}$
\citep[see also][]{Lopez2014}, where the scale height $H$ in the isothermal layer is, 
\begin{eqnarray}
{H\over{R_c}} &=& {{k_B T_{\rm eq}}\over{\mu m_H g R_c}} \nonumber \\ 
&\approx& 0.017\, \left({a\over{0.1{\rm ~AU}}}\right)^{-1/2}\left({{M_c}\over{5 {\rm ~M}_\oplus}}\right)^{-3/4}\left(\frac{R_p}{1.5\,{\rm R}_\oplus}\right)^2\, ,
\label{eq:Heq}
\end{eqnarray}
where we have estimated its value for a {\bc Sun-like host (mass 1~M$_\odot$, radius 1~R$_\odot$ and effective temperature 5780~K)} and for an envelope with a mean-molecular weight of $\mu = 2.35$ (solar composition), surrounding a core with an earth-like composition (see eq.~\ref{eq:rhocore}). 
%with $g$ the gravitational accleration and $a$ the orbital separation. Equation~\ref{eq:Heq} has been evaluated for a solar-type star and mean-molecular weight ($\mu$) of 2.35. 
%The radiative cap contains an envelope mass ($M_{\rm rad} \approx 4\pi R^2H \rho_{\rm rcb}$), $\lesssim 10^{-3}$\,M$_\oplus$ for the planets of interest. The planet's physical radius is set by the photospheric radius and this is what is determined in numerical models. Typically we find there is approximately six scale heights between the radiative-convective boundary and the photosphere Equation~\ref{eq:Heq} tells us that 
So the isothermal radiative cap is geometrically thin. In the following derivation, we take the planet radius to be approximately $R_p \approx R_{\rm rcb}$, but correct for the isothermal layer thickness in all relevant places (including all figures). 

\subsubsection{Convective interior}\label{sec:convective}
Adopting an equation of state for the adiabatic part of $\mathcal{P} = K \rho^\gamma$ with $\gamma$ and $K$ being constants,\footnote{In reality, for our ideal gas envelope $\gamma$ transitions from $\gamma=7/5$ in the upper envelope where molecules dominate to $\gamma=5/3$ in the lower envelope where molecules are dissociated. But the bulk of the planet can be considered as atomic.}
% not sure, depending on the degree of freedom...
hydrostatic equilibrium gives a density profile of: 
\begin{equation} \rho=\rho_{\rm rcb}\left[1+\nabla_{\rm ab}\left(\frac{GM_c}{c_s^2 R_p}\right)\left(\frac{R_p}{r}-1\right)\right]^{1/(\gamma-1)}\, ,\label{eq:den_profile} \end{equation} 
where $\nabla_{\rm ab} \equiv (\gamma-1)/\gamma$ is the adiabatic gradient, and the isothermal sound speed $c_s^2 \equiv \partial \mathcal{P}/\partial \rho|_T = \mathcal{P}/\rho$ is evaluated at $R_{\rm rcb}$.  For planetary atmospheres that are strongly bound ($v_{\rm esc}^2 \sim GM_c/R_p \gg c_s^2$)\footnote{Even if the initial planets may not be so, they rapidly evolve to such a state via the ``boil-off'' process investigated in \citet{OW16}.},  the unity term inside the braket can be safely ignored for much of the planetary interior that contributes significantly to the envelope mass, and we can simplify the above expression into \begin{equation} \rho\simeq\rho_{\rm rcb}\left[\nabla_{\rm ab}\left(\frac{GM_c}{c_s^2 R_p}\right)\left(\frac{R_p}{r}-1\right)\right]^{1/(\gamma-1)}\, .  \end{equation}
%This is convenient as it makes the integrals easier, and is easily verified as a good approximation when %$R_B/R_p\gtrsim 10$.
Solving for the mass enclosed in the atmosphere yields:
\begin{eqnarray}
M_{\rm env} & = & \int_{R_c}^{R_p}4\pi r^2\rho\, {\rm d}r \nonumber \\ 
& \simeq &
4\pi R_p^3\rho_{\rm rcb}\left(\nabla_{\rm ab}\frac{GM_c}{c_s^2R_p}\right)^{1/(\gamma-1)}I_2(R_c/R_p,\gamma) 
\label{eqn:Menv}
\end{eqnarray}
where the dimensionless integral $I_2$ is:
\begin{equation}
I_2(R_c/R_p,\gamma)=\int_{R_c/R_p}^1x^2\left(x^{-1}-1\right)^{1/(\gamma-1)}{\rm d}x\, .
\end{equation}
{\bc The properties of this dimensionless integral are discussed in the appendix}. For $\gamma=5/3$, as applies for the bulk of the atomic interior,
%(remember at such close separations the radiative cap has a temperature of $\sim 1000\,$K), 
the integrand for $I_2$ peaks at $x =1/4$. So in the limit of a puffy envelope, $R_c/R_p \leq 1/4$, $I_2$ is  fairly independent of $R_c/R_p$ and is of order unity. % I_1 ~ 0.15
While in the opposite limit of a thin envelope, quantified as $\Delta R = R_p-R_c \ll R_p$, the integral 
can be approximated as
%It is clear that for $\gamma=5/3$ that $I_1$ is not dominated at small radii, and therefore for %$R_c/R_p\ll1$, $I_1$ is approximately an order unity constant. However, for $\Delta R_p/R_c\ll1$, %$I_1$ is no longer constant and it must be evaluated. In this limit the integral $I_1$ becomes:
\begin{equation}
I_2 \approx  \nabla_{\rm ab} \left(\frac{\Delta R}{R_p}\right)^{\gamma/(\gamma-1)} \approx  \nabla_{\rm ab} \left(\frac{\Delta R}{R_c}\right)^{\gamma/(\gamma-1)}\, .
\end{equation}
With these expressions, the envelope mass fraction can be expressed as a function of the envelope thickness, $\Delta R/R_c$, {\bc in the two regions $\Delta R/ R_c < 1$ and $\Delta R/ R_c > 1$: } 
    \begin{eqnarray}
       X & \sim & 
            \frac{\rho_{\rm rcb}}{\rho_{\rm core}}
%\left(\frac{R_p}{R_c}\right)^3
\left(\frac{GM_c}{c_s^2R_c}\right)^{1/(\gamma-1)} \nonumber \\
& & \times         \begin{dcases}
\left(\frac{\Delta R}{R_c}\right)^{\gamma/(\gamma-1)} & \text{if } \Delta R/R_c <1  \\
\left(\frac{\Delta R}{R_c}\right)^{(3\gamma-4)/(\gamma-1)}
 & \text{if } \Delta R/R_c >1
        \end{dcases}\label{eqn:envelope_mass_frac}
    \end{eqnarray}
where we have dropped all order unity constants for clarity.

While the above expression relates $X$ to $\Delta R/R_c$, it still contains a variable $\rho_{\rm rcb}$ that depends on planet radius and envelope mass, among other things. To eliminate these dependencies, we appeal to the characteristics of the radiative-convective boundary. At this location, the temperature gradient remains adiabatic by definition, or ${{d\log T}/{d\log P}} = \nabla_{\rm ad}$, giving rise to,
\begin{equation}
{{d\log T}\over{dr}} = \nabla_{\rm ad} {{d\log P}\over{dr}} = - \nabla_{\rm ad} {{G M_c}\over{R_p^2 c_s^2}}\, .
\label{eq:hydrostatic}
\end{equation} 
The mode of energy transport changes at this point from advection by convective eddies to radiative diffusion, allowing us to write
\begin{equation}
\frac{d \log T}{d r} =- \frac{L}{4\pi R_p^2} \frac{3\kappa \rho_{\rm rcb}}{16\sigma T_{\rm eq}^4}\, ,
\label{eqn:rad_diff}
\end{equation}
where we have approximated values for the temperature and radius at the radiative-convective boundary as those at the surface, {\bc the impact of this approximation is discussed in \S~3.4}. The internal luminosity arises from gravitational contraction of the atmosphere, {\w i.e., changes in the gravitational binding energy $U$},\footnote{We ignore other internal heat sources, e.g., the heat capacity of the core and energy from radioactive decay. See discussion in \S~\ref{sec:core_lum}. }
\begin{equation}
{\bc L =\frac{{\rm d}U}{{\rm d}t} \approx \frac{U}{\tau_{\rm KH}}\approx  {1\over{\tau_{\rm KH}}}\left( \int_{R_c}^{R_p} {{GM_c \rho}\over{r}} 4\pi r^2 dr\right)\, . \label{eqn:L1}}
\end{equation}
{\w Using eqs. \refnew{eq:den_profile} \& \refnew{eqn:Menv}, we can re-write the luminosity as }
\begin{equation}
{\bc L \approx {1\over{\tau_{\rm KH}}}
{{GM_c M_{\rm env}}\over{R_p}} {{I_1(R_c/R_p)}\over{I_2(R_c/R_p)}}\, }, 
\label{eq:eqL}
\end{equation} 
where $\tau_{\rm KH}$ is the Kelvin-Helmholtz timescale (or cooling timescale). It is of order the planet age except in two cases: pre-cooling \citep[the case considered in][]{OW16} and high mass-loss rate (see \S \ref{subsec:complications}). $I_1$ is another  dimensionless integral given by
\begin{equation}
I_1 (R_c/R_p) = \int_{R_c/R_p}^1 x (x^{-1}-1)^{1/(\gamma-1)}\, {\rm d}x\, ,
\end{equation} 
{\bc whose relation to $I_2$ is discussed in the appendix. It is also shown there  that the ratio $I_1/I_2$ smoothly varies from $1$ to $\sim 3$ over the parameter range of interest}. Finally, we adopt an opacity law of $\kappa=\kappa_0 {\mathcal P}^\alpha T^\beta$ to obtain the following expression for the density at the radiative convective boundary, {\bc by substituting eqs. \refnew{eq:hydrostatic} \& \refnew{eq:eqL} into eq. \refnew{eqn:rad_diff}}, 
\begin{equation} \rho_{\rm rcb} \approx\left(\frac{\mu}{k_b}\right)\left[\left(\frac{I_2}{I_1}\right)\frac{64\pi\sigma T^{3-\alpha-\beta}_{\rm eq}R_p \tau_{\rm KH}}{3\kappa_0M_cX}\right]^{1/(1+\alpha)} 
  \label{eqn:rho_rcb} \end{equation}
Substituting this into eq. \refnew{eqn:envelope_mass_frac}, we obtain the final monotonic dependence of envelope mass fraction on planet radius of
 \begin{eqnarray}
       X  & \propto   & 
\left({{I_2}\over{I_1}}\right)^{n_I} \, \mu^{n_\mu}\, \kappa_0^{n_\kappa}\, 
T_{\rm eq}^{n_T}
\, \tau_{\rm KH}^{n_\tau} \, \rho_{M_\oplus}^{n_\rho}\, 
M_c^{n_{M}}\, \nonumber \\
& & \times       \begin{dcases}
            \left({{\Delta R}\over{R_c}}\right)^{n_a} \,\,\, & \text{if } \Delta R/R_c <1  \\
            \left({{\Delta R}\over{R_c}}\right)^{n_b}\,\,\, & \text{if } \Delta R/R_c >1
        \end{dcases}\label{eqn:X_R}
    \end{eqnarray}
where we have dropped all constants of physics and assumed that, due to compression, all solid cores (naked planets) have mass-radius relation as $M_c \propto R_c^4$ (\citealt{Lopez2014}, following, \citealt{Fortney2007}, \citealt{Valencia2010}), or
% sm.s/mass_radius, looks ok, rho_terrestrial=5.5, rho_ice=1.4, rho_iron=11
\begin{equation}
\rho_{c} = \rho_{M_\oplus} \left({{M_c}\over{1 M_\oplus}}\right)^{1/4}\, ,
\label{eq:rhocore}
\end{equation}
where $\rho_{M\oplus}$ is the density of a 1~M$_\oplus$ core and depends only on the core composition. For terrestrial composition, $\rho_{M\oplus}=5.5~{\rm g\,cm^{-3}}$, while it is $11$, $4$ and $1.4$~g$\,$cm$^{-3}$ for pure iron, silicate and water/ice cores respectively \citep{Fortney2007}.
 The power indexes are, respectively, 
\begin{eqnarray}
n_I & = & n_\tau = {1\over{\alpha+2}} \approx 0.37\, ,\nonumber \\  
n_\mu& = &  \left(1+{1\over{\gamma-1}}\right)\, {{\alpha+1}\over{\alpha+2}} \approx 1.57\, ,\nonumber \\
n_\kappa & = & - {1\over{\alpha+2}} \approx - 0.37 \, , \nonumber \\
n_T & = & \left({{{3-\alpha-\beta}\over{\alpha+1}}-{1\over{\gamma-1}}}\right)\, {{\alpha+1}\over{\alpha+2}} \approx -0.24\, ,\nonumber \\
n_{\rho} & = & - \left[ {1\over3}\left({1\over{\gamma-1}}-{1\over{\alpha+1}}\right)+1\right] \, {{\alpha+1}\over{\alpha+2}}\approx -0.82\, \, \nonumber \\
n_{M} & = & {2\over 3}
\left({1\over{\gamma-1}}-{1\over{\alpha+1}}\right)\, {{\alpha+1}\over{\alpha+2}} + {{n_\rho}\over 4} \approx 0.17\, ,\nonumber \\
n_a & = & \frac{\gamma(\alpha+1)}{(\gamma-1)(\alpha+2)} \approx 1.57\, \nonumber \\
n_b & = & \left({{3\gamma-4}\over{\gamma-1}}+{1\over{\alpha+1}}\right)\,{{\alpha+1}\over{\alpha+2}} \approx 1.31\, ,
\label{eq:Ns}
\end{eqnarray}
where we have also evaluated {these expressions} with $\gamma=5/3$, $\alpha=0.68$, $\beta=0.45$, the latter two identified by \citet{Rogers2010} as the opacity law appropriate for a solar metalicity H/He envelope of a low-mass, highly irradiated planet \citep[also see][]{Freedman2008}.  

So, at a given planet size ($\Delta R/R_c$), {\bc the exponents in eq.~\refnew{eq:Ns} tell us} the envelope mass is higher for a denser core composition (e.g., iron oxide vs. water ice), for an older planet, and for planets further away from their stars, all as expected.  Furthermore, the nature of the opacity law  makes the planet size a direct measure of the H/He envelope mass fraction, with weak sensitivities to all other factors. This has been noticed numerically in earlier works \citep[e.g.,][]{Lopez2014, Chen2016}. \citet{Lopez2014} provide  power-law fits to their numerical models over a wide range of parameter space, and the power-law indices listed in eq.~\refnew{eq:Ns} are similar to their results (eq.~4 in that paper, though theirs are fits for an ``enhanced opacity'' model, not solar-metallicity). A different opacity law may lead to very different results.
% This insensitivity is thus not something that naturally falls out of the basic physics, but is rather %sensitive to how the opacity varies with pressure and temperature unique to H/He dominated gas at the %temperatures and pressures appropriate for low-mass, close-in planets. 

Now, let us define an $X$ value at which the planet radius doubles, $\Delta R = R_c$, as $X_2$.  We shall argue later that $X_2$ is a crucial parameter for producing the evaporation valley. In eq. \refnew{eqn:X_R}, setting $\Delta R = R_c$ and re-inserting all constants that we have previously suppressed {\bc (with $\kappa=1.29\times10^{-2}$~cm$^{2}$~g$^{-1}$ at a pressure of 1 bar and a temperature of 1000~K, \citealt{Rogers2010})}, and for solar metallicity gas, we find:
%\begin{equation}
%X_2\approx 0.03\left(\frac{M_c}{6.5\,{\rm M}_\oplus}\right)^{\frac{5\alpha-1}{8(\alpha+2)}}\left(\frac{a}{0.1\,{\rm AU}}\right)^{-\frac{3-4\alpha-\beta}{4{2+\alpha}}}\left(\frac{\tau_{\rm KH}}{100\,{\rm Myr}}\right)^{\frac{1}{2+\alpha}}\left(\frac{\mu_{\rm rcb}}{2.35}\right)^{\frac{5(1+\alpha)}{2(2+\alpha)}}\label{eqn:X2_1}
%\end{equation}
\begin{eqnarray}
X_2 & \approx & 0.027\, 
\left(\frac{P}{10\,{\rm days}}\right)^{0.08}\, 
\left(\frac{M_*}{M_\odot}\right)^{-0.15}\, 
\left(\frac{\tau_{\rm KH}}{100\,{\rm Myr}}\right)^{0.37} \, \nonumber \\
& \times & 
\left({{\rho_{M_\oplus}}\over{5.5\, {\rm g\,cm^{-3}}}}\right)^{-0.82}\, 
\left(\frac{M_c}{5\,{\rm M}_\oplus}\right)^{0.17}\, .
\label{eq:X2value}
%\left(\frac{\mu_{\rm rcb}}{2.35}\right)^{1.6}
\end{eqnarray}
where $P$ is the orbital period. We have also set $T_{\rm eq}= (L_*/16\pi a^2)^{1/4}$ and adopted an empirical mass-luminosity relation for low-mass dwarfs: $L_*/L_\odot = (M_*/M_\odot)^{3.2}$ {\w \citep[see, e.g.][]{Allens}}.  In principle, the value of $I_1/I_2$ also depends on $\Delta R/R_c$. But since it varies by only a factor of $3$ within our range of interest, together with the weak index $n_I$, this can be safely ignored. { Moreover, when the thickness of the isothermal layer is accounted for, the true $X_2$ is slightly smaller. We actually obtain that $X_2$ is of order a few percent at an age of a Gyrs.}

In summary, the envelope mass fraction, at which the planet's radius doubles, is of order a few percent, and is a relatively weak function of planet period, stellar mass, core mass and planet age.
%We have constructed polytropic planet models that confirm the validity of equation \refnew{eqn:X_R}.

\subsection{Timescale for atmospheric erosion}
%-- analytics and numerical}
We define a mass-loss timescale for envelope evaporation as
\begin{equation}
t_{\rm \dot{X}}\equiv \frac{X}{\dot X} = \frac{M_{\rm env}}{\dot{M}_{\rm env}}\, .\label{eq:tx}
\end{equation}
In the following, we show that this timescale {\it naturally} peaks for envelopes with masses $X \approx X_2$, or those which double the core's radius. 
% Equation~\ref{eqn:X_R} implies that unless the opacity falls with very rapidly with pressure then the %mass-loss timescale {\it will always} peak around the envelope mass fraction at which the radius  %doubles.

The mass-loss rate is given by the ratio of photo-evaporative power and the binding energy of the planet \citep{Lecavelier2007,Erkaev2007}. 
Let $L_{\rm HE}$ be the luminosity of high energy photons from the star, and 
the dimensionless factor $\eta$ be the efficiency of these photons for mass-removal,\footnote{This efficiency is defined as if the high-energy photons are intersected by a cross-section $\pi R_p^2$. The definition of this efficiency is arbitrary and many other definitions do exist in the literature.}
 the mass-loss rate is 
\begin{equation}
\dot{M}_{\rm env}=\eta\frac{\pi R_p^3 L_{\rm HE}}{4\pi a^2 GM_p}\, .
\label{eqn:cond2}
\end{equation}
 The efficiency factor is not necessarily a constant, as has been demonstrated in multiple works \citep[e.g.][]{MurrayClay2009,Owen2012,Shematovich2014,OA16,
Salz2016}.  However, when compared to full radiation-hydrodynamic models it takes a value of order $0.1$ for low-mass planets \citep{Owen2012,OA16}. 
%\citep[see, e.g.][]{Lecavelier2007,Erkaev2007}:
For simplicity, we adopt a constant $\eta=0.1$ \citep[the so-called ``energy-limited'' approach, see e.g.,][]{Lopez2013}. However, we discuss its variation and impact on the planet population in \S \ref{subsec:model2}.  {\bcc We do not consider an ``effective absorption radius'' or ``expansion radius'' \citep[e.g.][]{Baraffe2004} that accounts for the higher radius at which the high-energy flux is absorbed compared to the planet's radius, as this effect can always be folded into the efficiency factor. Such a radius is also difficult to define for X-ray driven evaporation (which is important for low-mass planets) as different wavelength photons can be absorbed at very different radii.  We also neglect the effect of stellar tides \citep[e.g.][]{Erkaev2007} since most observed planets are far from their Roche radii. Specifically $<13\%$ of the planets (taking them to have a mass of 3~M$_\oplus$) used in the \citet{Fulton2017} sample would have their mass-loss rates increased by $>20\%$ and $<2\%$ would have their mass-loss rates increased by a factor of $>2$  using the \citep{Erkaev2007} prescription.}

Substituting eq. \refnew{eqn:X_R} into the above expressions, we obtain: 
\begin{eqnarray}
t_{\dot X} & \approx & {\rm 210 Myrs} \, 
 \left({{\eta}\over{0.1}}\right)^{-1} \, 
\left({{L_{\rm HE}}\over{10^{-3.5}L_\odot }}\right)^{-1} \,
\left(\frac{P}{10\, {\rm days}}\right)^{1.41}\, \nonumber \\
& & \times 
\left(\frac{M_*}{M_\odot}\right)^{0.52}\, 
\left({f\over 1.2}\right)^{-3} \, 
\left(\frac{\tau_{\rm KH}}{100\,{\rm Myr}}\right)^{0.37} \, \nonumber \\
& & \times \left({{\rho_{M_\oplus}}\over{5.5\, {\rm g\,cm^{-3}}}}\right)^{0.18}\, 
\left(\frac{M_c}{5\,{\rm M}_\oplus}\right)^{1.42} \, 
\nonumber \\
& & \times       \begin{dcases}
            \left({{\Delta R}\over{R_c}}\right)^{1.57} \,\,\, & \text{if } \Delta R/R_c <1  \\
            \left({{\Delta R}\over{R_c}}\right)^{-1.69}\,\,\, & \text{if } \Delta R/R_c >1
        \end{dcases}
\label{eq:tdotx}
\end{eqnarray}
where we evaluate for solar-metallicity gas, and have adopted a parameter $f$ to account for the difference between $R_{\rm rcb}$ and the photospheric radius, $R_p = f R_{\rm rcb}$. { In our model, $f$ is computed self-consistently, by locating the photosphere at where ${\mathcal P} = (2/3) g/\kappa$.
%The largely isothermal atmosphere above it contains a non-negligible number of scale heights. 
The photosphere radius is then at $nH $ above the radiative-convective boundary, with $H$ being the local scale height, and $n = \ln (\rho_{\rm rcb}/\rho_{\rm ph})$, {\bc where $\rho_{\rm ph}$ is the density at the photosphere}. }
% here the definition of $R_p$ does include the added effect of the planet's radiative zone where its %height is determined by solving Equation~\ref{eq:Heq}, and its impact is parametrised by the order %unity factor $f$.

According to {eq. (\ref{eq:tdotx})}, the mass-loss timescale reaches a maximum at the point at which the radius is doubled ($\Delta R \approx R_c$, $X=X_2$). 
%We define this peak evaporation timescale as $t_{\rm peak} = t_{\dot X} (X=X_2)$.  
Such a behaviour arises because, below $X_2$, the planet's radius is dominated by that of the core and is independent of the envelope mass. So the photoevaporation timescale decreases for lower envelopes. This is an unstable situation that can cause complete stripping.
% If the mass-loss timescale decreases with decreasing envelope mass fraction (as it does when $X<X_2$ %then the entire atmosphere is unstable and will be completely stripped. Alternatively if the mass-loss %timescale increases with decreasing envelope mass fraction (as it does when $X>X_2$) then the planet %will only lose mass until the mass-loss time-scale becomes comparable to its cooling timescale again.
While above $X_2$, increasing $X$ leads to a radius expansion. The expansion is so significant that, for our adopted opacity laws and a constant $\eta$, the evaporation timescale shortens. {\bc This behaviour can be shown to occur for all well known evaporation models: energy-limited evaporation, when the radius increases faster than $X^{1/3}$; photon-limited evaporation \citep{OA16}, where the radius increases faster than $X^{1/2}$; and recombination limited evaporation \citep{MurrayClay2009}, where the mass-loss rate is exponentially sensitive to radius \citep{OA16} such that the erosion time-scale always decreases with increasing envelope mass fraction.} The fact that the envelope-mass loss time scale peaks for low-mass planets with envelope mass fractions of order a few percent was also noticed numerically by \citet{Chen2016} using {\sc mesa} models. With our {\bc minimal} model, it becomes obvious that the mass-loss timescale always peaks around $X = X_2$. For this feature not to happen, one requires $n_b<3$, or in terms of the opacity scaling with pressure, $-7/3<\alpha<-2$, which is unrealistic.

The heuristically derived eq. \refnew{eq:tdotx} has a sharp discontinuity at $\Delta R/R_c=1$. Numerically, we smooth the transition in the preceding calculations by using the fact, in reality $R_p=R_c+\Delta R$, rather than the approximation $R_p=R_c$ for $\Delta R < R_c$ and $R_p=\Delta R$ for $\Delta R > R_c$, presented above.    

\begin{figure} \centering \includegraphics[width=\columnwidth]{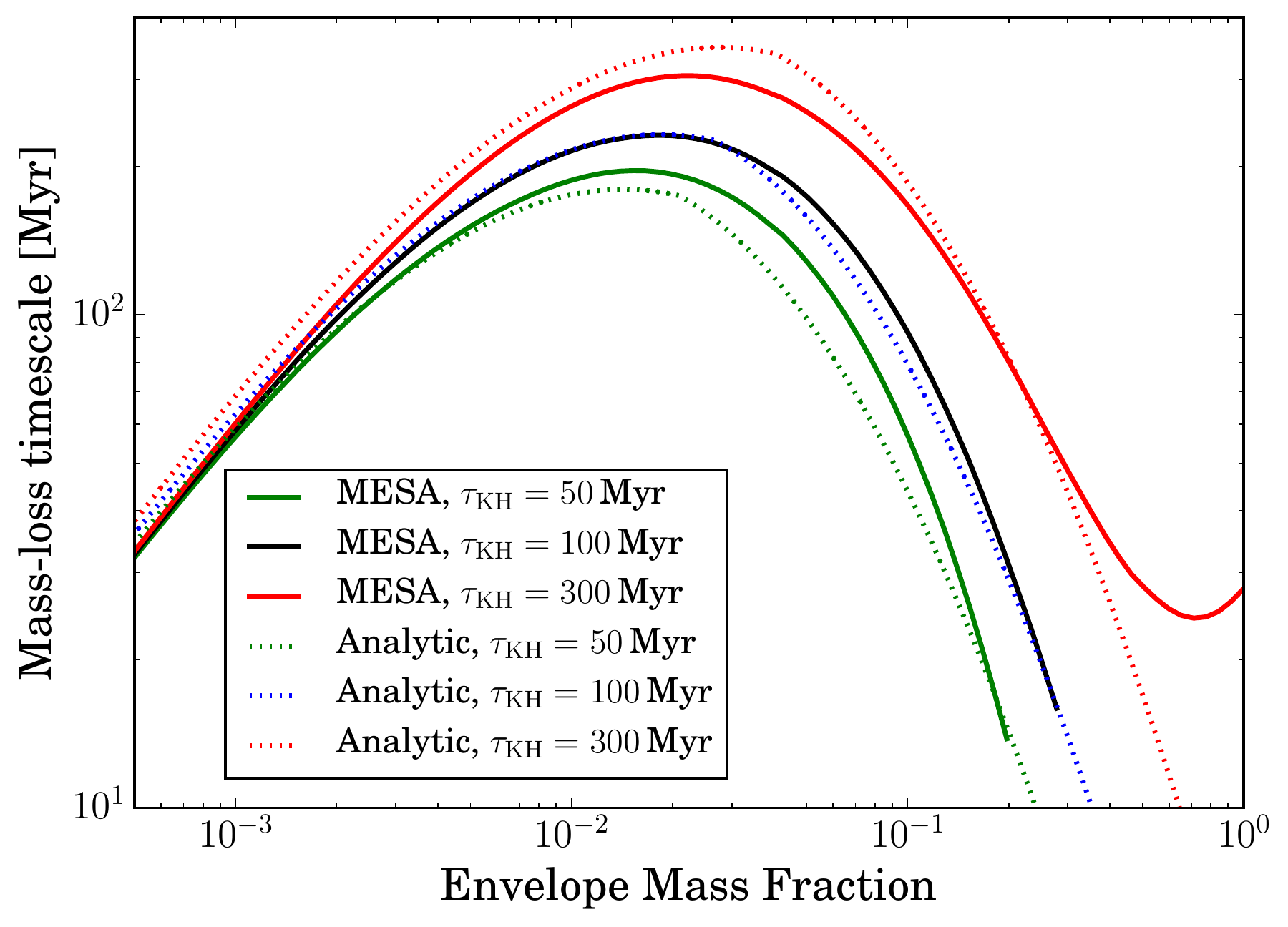} \caption{ The timescale for envelope evaporation is plotted as a function of envelope mass fraction, from
 numerical models calculated with the {\sc mesa} code (solid lines) and from our analytical model (eq. \ref{eq:tdotx}, dotted curves, smoothed as described in the text).  
The planet has a core mass of $5 M_\oplus$, an Earth-like core composition,
%($\rho_{\rm core}\approx 8\,$g~cm$^{-3}$) 
and lies at a period of 10~days around a {\bc "Sun-like"} star. See text for our choice of other parameters. 
%We assume energy-limited evaporation to calculate the mass-loss timescale with an %efficiency of 0.1. 
Three types of models with different cooling ages are plotted.
%include the radiative zone included in the definition of the planet's radius. Note the %numerical model's mass-loss timescales begin to increase with envelope mass fraction
The numerical model with the longest cooling age shows an uptick at large envelope mass, resulting from self-gravity of the envelope, {\bcc which compresses the envelope},  an effect we ignore in our analytical model. For models with shorter cooling ages, numerical models at these masses do not converge, as their radii typically exceed the planet's Bondi radius.} 

%{\w I DON''T AGREE WITH REASON GIVEN IN ORIGINAL TEXT. ``as their radii become so large the envelope becomes thermally unbound.'' ONE SHOULD BE ABLE TO FIND INITIAL STATES FOR ALL MODELS, NO? DELETED IT FOR NOW.}
%The numerical models show an uptick $X\gtrsim 0.5$) as self-gravity of the envelope %becomes important and the planet's radius becomes insensitive to changes in the %envelope mass fraction. Numerical models for envelope mass fractions $\gtrsim 0.3$ cannot be computed for the shortest cooling times (50 and 100~Myr) as their radii become so large the envelope becomes thermally unbound. 
\label{fig:mesa_compare} \end{figure}
In the following, we will construct {\bc planetary structure} models, based on eqs. \refnew{eqn:Menv}, \refnew{eq:eqL} and \refnew{eqn:rho_rcb}.  In Fig. \ref{fig:mesa_compare}, we compare results from our {\bc minimal} models against those produced using the {\it mesa} code \citep{Paxton2011,Paxton2013,Paxton2015}, suitably modified for highly irradiated low-mass planets \citep[see][]{Owen2013,Owen2016}. The agreement between the {\bc analytic} model and the {\sc mesa} results is good. The mass-loss timescale peaks at $X \approx 0.02-0.03$, roughly where the planet doubles in size, {\bc increases slowly with} cooling time; and the mass-loss timescale falls off as predicted on both sides of the peak.  The agreement only fails for cases where $X \gtrsim 1$, where the self-gravity of the envelope (ignored in our analytical models) becomes dominant. At large envelope masses, self-gravity keeps the planet's radius roughly constant to of order a Jupiter radius, and the mass-loss time begins to increase again as expected (it is well known hot jupiters are stable to evaporation, e.g. \citealt{Yelle2004}).

\section{Comparing to the observations -- the ``evaporation valley''}

Based on refined stellar parameters from the CKS survey, \citet[][]{Fulton2017} reported a ``gap'' in the planetary radius distribution \citep[also noticed by][based on the cruder KIC data]{Owen2013}: a deficit of planets at radii $1.8 \pm 0.2 R_\oplus$ (``gap'') that appears to extend from 10 days to 100 days in orbital period (``valley'').  We now use our analytical model to calculate the exact valley location for different parameters.
In the following, we detail our model choices and present results.

We make no effort to ``fit'' the observations at this stage, especially with our {\bc minimal} model. That is best left to comparisons with numerical models. Nevertheless, as we will see, even our model can be used to make strong inferences about the properties and possible formation channels of the {\it Kepler} planets. Usefully, we can analytically understand the origin of these constraints.

\subsection{The Integration}

\begin{figure} \centering \includegraphics[width=0.45\textwidth,trim=40 190 20 100,clip=]{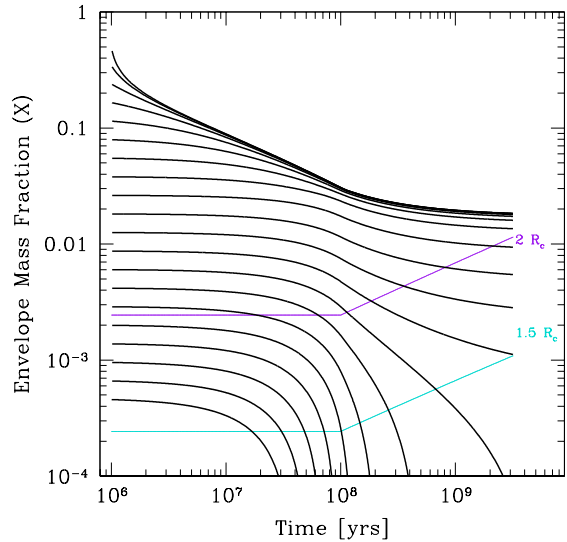} \caption{The erosion of atmosphere as a function of time, for planet models with a range of initial envelopes. All parameters are the same as in Fig.~\ref{fig:mesa_compare}. 
%but share the same parameters otherwise: solar type star, $L_{\rm sat}=10^{-3.5} %L_\odot$, $P=10$ days, $M_c = 5 M_\oplus$, $\eta=0.1$, $\rho_{M_\oplus}=5.5 {\rm %g/cm^3}$, $R_c = 1.3 R_\oplus$ (scaled values as in eqs. \ref{eq:X2value} \& %\ref{eq:tdotx}) and adopt an X-ray decay law of $a_0=0.5$ (eq. \ref{eq:LHE}).  
As expected, almost all erosion occurs in the first 100 Myrs, when the planets are hot and when the stars are bright in high-energy radiation. Low mass envelopes are stripped clean, while higher mass ones are herded towards $X \sim 1\%$ (and $R_p \sim 2 R_c$). The colored lines denote planet radii (values as marked, {they differ from the estimate in eq. \refnew{eq:X2value} because these are photospheric radius}).
%and are corrected as in \S  \ref{subsec:complications}.
This set of models resemble group (c) in Fig. \ref{fig:diagram}.
%The heavy line shows the planet model that has lost $50\%$ of its initial envelope. -- incorrect
% almost all models do
 %producing a deficit of planets at radii between $R_c$ and $\sim 1.5 R_c$. 
%The right side panel shows the histograms for an assumed initial (dotted) and final (solid)  distributions.
 }\label{fig:xevolve} \end{figure}

Starting from an initial envelope mass fraction at time zero, which we set to be 1~Myr \citep[roughly the disk dispersal timescale, e.g.][]{Hernandez2007}, we evolve the envelope mass fraction according to:
\begin{equation}
\frac{{\rm d}X}{{\rm d}t}=-\frac{X}{t_{\dot{X}}}\label{eq:evolve}\, .
\end{equation}
according to eq. \refnew{eq:tdotx}, and a prescription for the stellar high energy luminosity (\S \ref{sec:star_parameters}),  and the cooling of the planet (\S \ref{subsec:planetensemble}). 
Some example evolutionary tracks for a planet with different initial envelope mass fractions is shown in Fig.~\ref{fig:xevolve}.

%\subsection{Accuracy of the simple model}
We have also applied our analytical model to the envelope evolution of the {\it Kepler}-36b/c system and compare them against the more detailed {\it mesa} calculations presented in \citet{Owen2016}. 
We find that while we can reproduce their general results, namely, the lower mass {\it Kepler}-36b is completely stripped off, while the higher mass {\it Kepler}-36c retains a bulky envelope, our models suffer moderately more  mass loss for the latter planet during the early stages when its mass-loss rate is high, due to our assumption of constant mass-loss efficiency, an assumption not adopted by \citet{Owen2016}.
%As a result, the more massive planet can only retain about half the envelope mass as %it is observed today.

\subsection{Stellar Parameters}
\label{sec:star_parameters}

{We adopt host star masses similar to those in the CKS sample \citep{Fulton2017}, a Gaussian distribution in mass, centred at $1.3 M_\odot$ with a variance of $0.3 M_\odot$.}
%\begin{equation}
%   M_*:   \text{Gaussian}, <M_*>= 1.3 M_\odot, \sigma_M= 0.3 M_\odot \, 
%\label{eq:Mstar|
%\end{equation}

For the magnitude and the evolution of the high-energy flux ($L_{\rm HE}$, including UV through X-ray radiation), we adopt the empirical relation for main-sequence dwarfs, as summarized by  \citet{Jackson2012}
\begin{eqnarray}
  L_{\rm HE} = \begin{dcases}
 {L_{\rm sat}} &  {\rm for}\,\,   t < t_{\rm sat} \, \\
  L_{\rm sat} \left({t\over{t_{\rm sat}}}\right)^{-1-a_0}  & {\rm for}\,\, t \geq t_{\rm sat} \,.
\end{dcases} 
\label{eq:LHE} \end{eqnarray} 
We further choose $a_0=0.5$ and $t_{\rm sat}=100$ Myrs, and 
$L_{\rm sat} \approx 10^{-3.5} L_\odot (M_*/M_\odot)$,  motivated by 
the body of observational and modelling works \citep[e.g.][]{Gudel1997,Ribas2005,Jackson2012,Tu2015}.
Since $a_0>0$, the time-integrated high-energy ``exposure'' \citep{Lecavelier2007,Jackson2012,Owen2013} is dominated by that in the first $100$ Myrs. So the exact choice of $a_0$ has  little bearing on the final planet properties.
% rather the planets properties after billions of years of evolution are 
%imprinted during the first 100~Myrs of evolution \citep{Owen2013}.  

\subsection{The Planet Ensemble}
\label{subsec:planetensemble}

We now must decide what the primordial {\it Kepler} planets look like.  In this work, we consider only one population of planets -- we discuss the evidence for a second population in \S \ref{sec:two_pop}.

We adopt the following orbital period distribution for planets around all stars, 
\begin{eqnarray}
 {{dN}\over{d\log P}} \propto \begin{dcases} 
 {\rm constant} &  \text{for} \, P > 7.6\, \text{days} \\
  P^{1.9} & \text{for}\, P \leq 7.6\, \text{days} \, .
\end{dcases}
\label{eq:2groupb}
\end{eqnarray}
This distribution is obtained by fitting the {\it Kepler} planet sample and correcting for transit probabilities. It is similar to those obtained in earlier work \citep[e.g.][]{Fressin2013}.
%motivated by the observed {\it Kepler} sample \citep{Petigura,??,Silburt} as well as 
% {\w see sm.s/read13}.  

For the planetary cores, we assume them to be Earth-like ($\rho_{M_\oplus} = 5.5\, {\rm g~cm^{-3}}$). In \S \ref{sec:core_composition}, we will vary this parameter and show that the observed CKS sample actually demands this choice.  Furthermore, we take a Rayleigh distribution for the core mass
\begin{equation}
 {{dN}\over{dM_c}}  \propto M_c \exp^{- {{M_c^2}/{2\sigma_{M}^2}}} \,. 
% \, \nonumber \\
%& {\rm Pop\, II} & \nonumber \\
%&{{dN}\over{dM_c}} & \propto M_c \exp^{- {{M_c^2}/{2\sigma_{MII}^2}}} \, \nonumber \\
%&  X_0 & = 0 \, .\nonumber \\
\label{eq:2group}
\end{equation}
The actual mass distribution of {\it Kepler} planets has not been reliably established. {Early RV studies \citep[e.g.][]{Howard2010,Mayor2011} showed that the planet occurrence rate increased towards low-masses: planets in the mass range 3-10~M$_\oplus$ being most common, with the occurrence rate falling rapidly towards higher masses.} One {more recent} attempt is provided by \citet{Marcy2014} where they selected 22 KOIs and measured masses (or their upper-limits) for 42 planets.
%Planet candidates with radii between $2-5R_\oplus$ and with likely detectable RV %signals are preferrentially observed. 
The mass function for this group can roughly be fit as a Rayleigh distribution with a mode $\sigma_M \sim 5\pm 1\,$M$_\oplus$.
%sm.s/readmarcy, none is a good fit, there seems to be two lumps, one lump with 
% sigma_M=2, the other with sigma_M=5
However, this sample is arguably biased towards higher masses -- compared to {\it Kepler} planets within the same period range (say, $< 20$ days), this sample contains planets that are, on average, larger and therefore likely more massive. 
% see Fig. 53 of Marcy 2014. from 1 to 3 R_E within 20 days, roughly 
% sm.s/readmarcy
%Another issue is that the 22 KOIs have a mean mass of $0.97 M_\odot$, lower than the CKS sample. 
We therefore choose a slightly smaller mass scale of $\sigma_M=3\,$M$_\oplus$. We are also motivated by the position of the small-size peak. At a radius of $1.3\,$R$_\oplus$, this corresponds to a terrestrial planet with $M_c=3\,$M$_\oplus$.

We also need to prescribe the initial envelopes.  This includes both their initial thermal time (which determines the initial entropy) and initial masses. \citet{OW16} argued that after low-mass planets are born, they undergo a rapid phase of cooling and mass-loss (the ``boil-off'' phase) and age prematurely to $\tau_{\rm KH} \sim 100$ Myrs within a relatively short time.  After about 100Myrs, they continue to cool off normally. So we take 
\begin{eqnarray}
  \tau_{\rm KH} = \begin{dcases}
 10^8 ~{\rm yrs}\, & {\rm for}\,\, t < 10^8 ~{\rm yrs}\,  \\
  t   & {\rm for}\,\,  t \geq 10^8 ~{\rm yrs} \, .
\end{dcases}
\label{eq:tauKH}
\end{eqnarray} 
Due to the fact $a_0> 0$ and the above form of cooling contraction, most of the envelope erosion occurs in the first 100 Myrs, and there is little change to the planet atmosphere after that. 
%A choice of $\tau_{\rm kh}=t$ at all times would only the first 100 Myrs more important; however we %emphasis that even for a cooling planet it is the mass-loss that occurs around 100 Myrs that matters %\citep{Owen2013}.

\begin{figure} \centering \includegraphics[width=0.48\textwidth,trim=20 170 10 110,clip=]{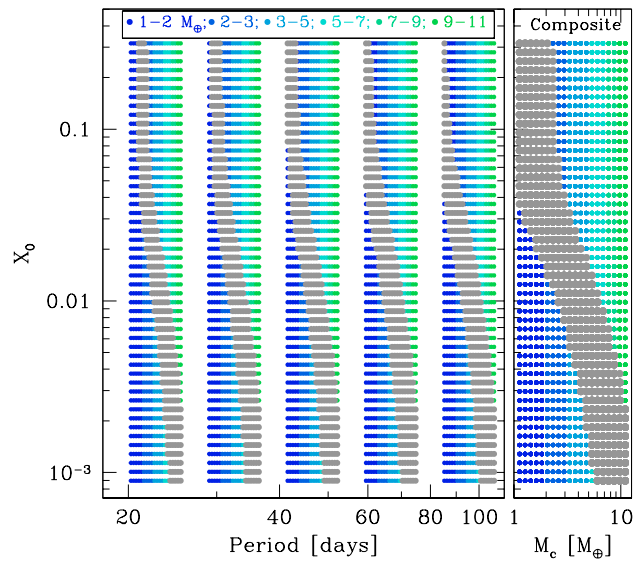} \caption{For a range of initial envelope mass fractions (y-axis), orbital periods (x-axis) and core masses (represented by different colors, slightly dispersed to the right for clarity), we calculate the final planet radius, at a few Gyrs, due to both cooling contraction and photoevaporation. Models are marked in grey if the final radii fall within the observed gap ($R = 1.8 \pm 0.2 R_\oplus$) and are thus disfavoured by the observations. The left panel shows the period dependence of these models, while the right panel shows the ``composite'' view: grey indicate any models that are excluded within the range of $P=20-100$ days (the period range that the gap is clearly visible in data, \citealt{Fulton2017}). The initial planet population that can satisfy the observation fall into two categories: planets with masses more than two Earth masses and initial H/He masses more than a few percent; or lower mass planets with essentially no atmosphere.  The star is assumed to be sun-like but the results are not {\bc particularily} sensitive to the stellar mass.  }\label{fig:xgap} \end{figure}

To guide our choice for the initial envelope mass, we first ask what kind of planets would have occupied the observed gap, at an age of a few billion years. These are the grey points in Fig. \ref{fig:xgap}, for cores of terrestrial composition ($\rho_{M\oplus}= 5.5 ~{\rm g~cm^{-3}}$), with a range of core mass, initial envelope mass, and orbital period. Since these models would appear in the CKS gap, they are disfavoured by the observations. Now we make the important assumption that planets are born with the same envelope mass and core mass distributions across the range of period of interest (from a few to $100$ days). This assumption mostly reflects our ignorance on the proper initial condition, and is almost certainly not true in reality. But under this restrictive assumption, all models that are marked with grey in the right-hand panel of Fig. \ref{fig:xgap} are not permitted. In other words, there are two groups of planets that are likely progenitors for the observed {\it Kepler} planets: one group are planets that are slightly more massive ($M_c > 3$~M$_\oplus$) and are born with at least a few percent of H/He; the other group, planets that are less massive ($M_c < 3$~M$_\oplus$) and that are born, for all purposes, bare. The second group cannot account for planets with radii larger than $1.8$~R$_\oplus$. { Furthermore, \citet{Fulton2017} indicate that the completeness of their exoplanet radius distribution becomes uncertain at small planet radii $<1.14$~R$_\oplus$. Inspection of Fig \ref{fig:xgap} indicates that planets with masses $<2$~M$_\oplus$ ($R_p=1.18$~R$_\oplus$) are disfavoured if they contain H/He envelopes with masses of a few percent or more. Therefore, while the current CKS sample cannot determine if planet occurrence drops significantly below 1.14~R$_\oplus$ we predict it should, unless there is a second population of planets (see \S\ref{sec:two_pop})}. {\w Moreover, we emphasize that the choice of the mass scale, $\sigma_M=3\,$M$_\oplus$, is motivated by both the radii of the bare planets, {\bf and} the size distribution of planets to the right of the valley.}

Guided by this insight, we adopt a logarithmically flat distribution for $X_0$, the initial envelope fraction, with $X_0 \in [X_{\rm min}, X_{\rm max}]$ and $X_{\rm min} = 0.01$ and $X_{\rm max} = 0.3$.
% {{dN}\over{d\log {X_0}}} & =  {\rm constant} \hskip0.5in  X_0 \in [X_{\rm min}, X_{\rm max}] \, .
% when $\sigma_{M}= 3 M_\oplus$, the best fit $X_{\rm min} = 0.03$; while when $\sigma_{M} =6 M_\oplus$, the best fit is $X_{\rm min} = 0.005$. {\w This can be seen from eq. \refnew{eq:???}.} However, the latter values of $\sigma_{M}$ is inconsistent with the upper envelope of radius as a function of period -- observations show that most planets inward of $10$ days are naked cores. So we prefer a value of $\sigma_{M} = 3 M_\oplus$ and $X_{\rm min} = 0.03$. {\w what about an even smaller $\sigma_{MI}$?}
The value of $X_{\rm max}$ matters little as long as it lies well above $1\%$, while the value of $X_{\rm min}$ is suggested by Fig. \ref{fig:xgap}, for our choice of core masses.
%If we have adopted a larger $\sigma_M$, the value of $X_{\rm min}$ can lie lower.

\subsection{A few details}
\label{subsec:complications}

The radius measured for a planet {by the} transit {method} is not the photospheric radius (which accounts for the radial light-path for the local black-body photons), but should be determined by the tangent light-path for transmitting stellar photons. The chord is longer for the latter group by a factor of $\sim \sqrt{8 R/H}$, increasing the transit radius by $\sim \ln (\sqrt{8 R/H}) H\sim 3 H$ {\bc \citep[e.g.][]{Lopez2014}}.  However, opacity for optical photons is smaller than that for infra-red photons, leading to a ``deeper photosphere''. These two effects cancel each other to some degree and we decide to ignore them here.

%but should be several scale heights larger due to the large atmospheric column a %transit chord intersects. Instead of one scale height for vertically travelling light, the %path length is now $\sim 2 \sqrt{(R+H)^2 - R^2} \sim \sqrt{8 R H}$. The corresponding %transmission photosphere is at $\rho_{\rm ph,tr} = \rho_{\rm ph} \sqrt{H/8R} \sim %0.05 \rho_{\rm ph}$, or about $n \approx 3$ scale heights up \citep[see %also,][]{Rogers2010,Lopez2014}. Thus the observed transit radius  should be $nH$ %larger than the photospheric radius we discuss above. 

Another complication arises from how we treat the thermal evolution of the planet in presence of mass-loss. Here, we simply assume the two are independent (eq. \ref{eq:tauKH}). This is inappropriate when the envelope is being evaporated faster than cooling contraction. In this case, the lifting of pressure at the top allows the envelope to expand adiabatically. The irradiated atmosphere then actually transports heat inward, maintaining the same internal entropy.  The radiative-convective boundary remains fixed at the same density, and the internal entropy can be transported out with the same cooling luminosity as before. This accelerates the cooling compared to the case of no-cooling (this is the physical basis for the ``boil-off'' discussed in \citealt{OW16}, see also the discussion in \citealt{Ginzburg2016}). 
%But more importantly, the envelope appears larger than the same mass at the same %age. This accelerates mass-loss and leads to divergent behaviour. In the following, %we set $\rho_{\rm rcb} = constant$ once this event occurs. 
We do not correct for this effect in our {\bc minimal} model, but note that this may lead to enhanced mass loss in some cases; however, models that typically enter this region are on their way to be completely stripped anyway. 

\subsection{Planets at a few Gyrs}

%\subsection{Resulting population after billions of years}

\begin{figure} 
{\includegraphics[width=5.92cm,trim=90 195 65 120,clip=]{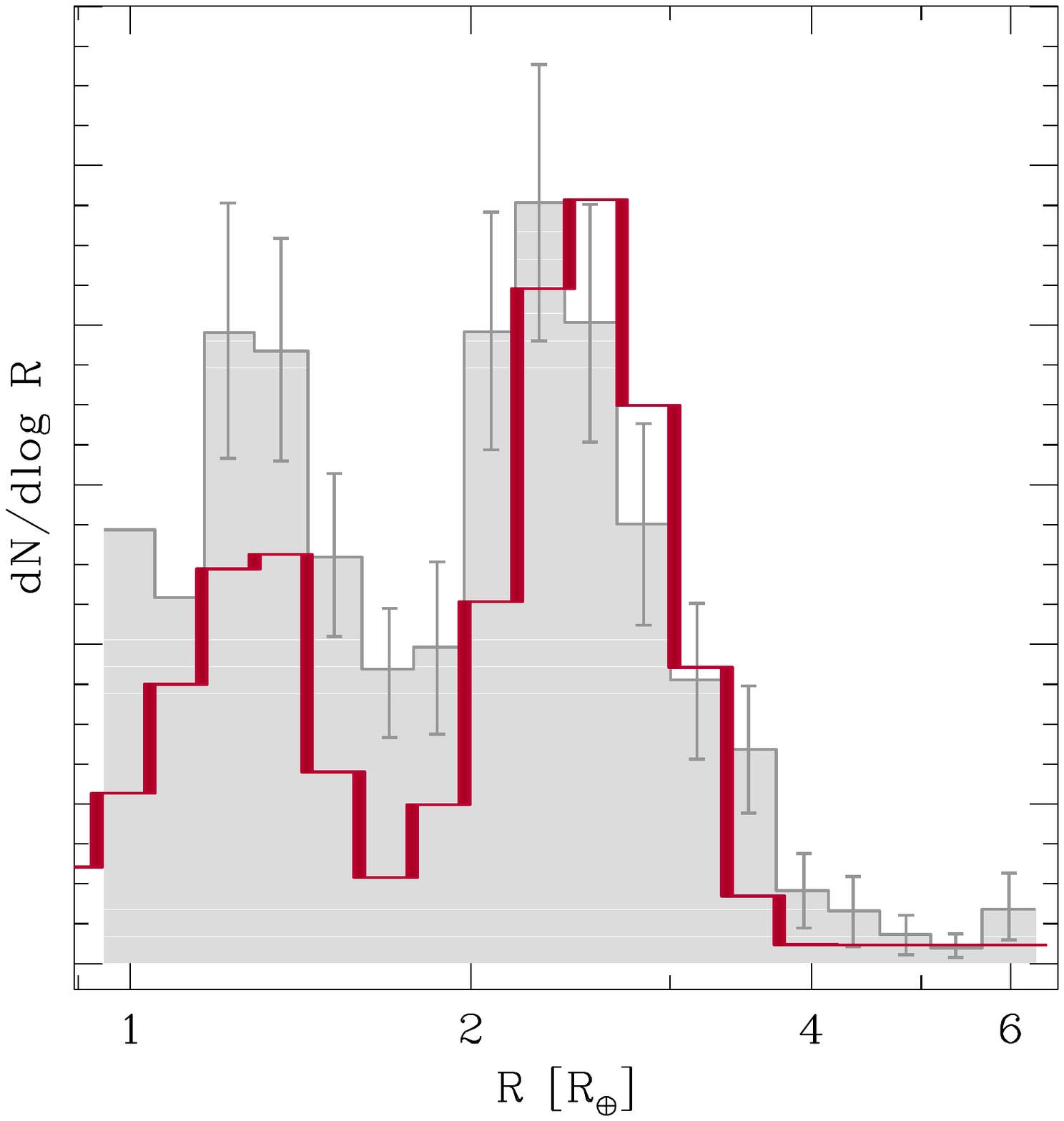}}
{\includegraphics[width=2.58cm,trim=105 175 300 120,clip=]{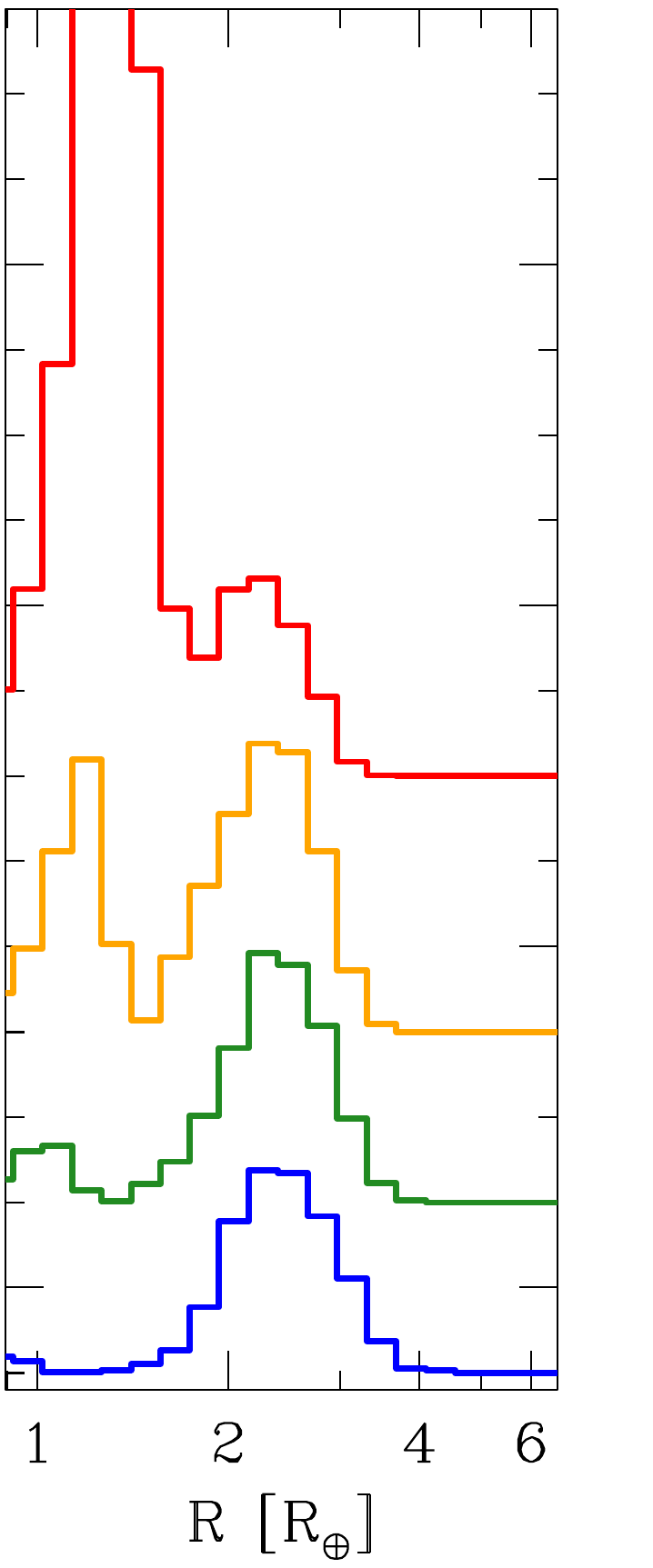}}
\caption{The final radius distribution after three billion years of evolution. The left panel shows the distribution for our entire model population (colored line). It compares well against the observed one from \citep{Fulton2017}, plotted here as shaded grey histogram,
%a Rayleigh distribution of core masses with $\sigma_M = 5 M_\oplus$ and an initial %logarithmically flat spread in $X \in [10^{-2},0.3]$, terrestrial composition. 
  including the gap at $1.8 R_\oplus$, the peaks at $1.3 R_\oplus$ and $2.6 R_\oplus$, as well as the sharp deficit of Neptune-like planets (size beyond $3 R_\oplus$), note we did not attempt to ``fit'' the observed histogram (see text). The right panel displays the radius distribuion binned by orbital periods. From top to bottom, the period bins are, $0-10$, $10-20$, $20-40$, and $40-100$ days. }\label{fig:xvalley_rayleigh}
\end{figure}

\begin{figure*} 
\centering
%{\includegraphics[width=9.6cm,trim=120 760 65 200,clip=]{x_rayleigh.png}}
{\includegraphics[width=8.5cm,trim=0 0 0 0,clip=]{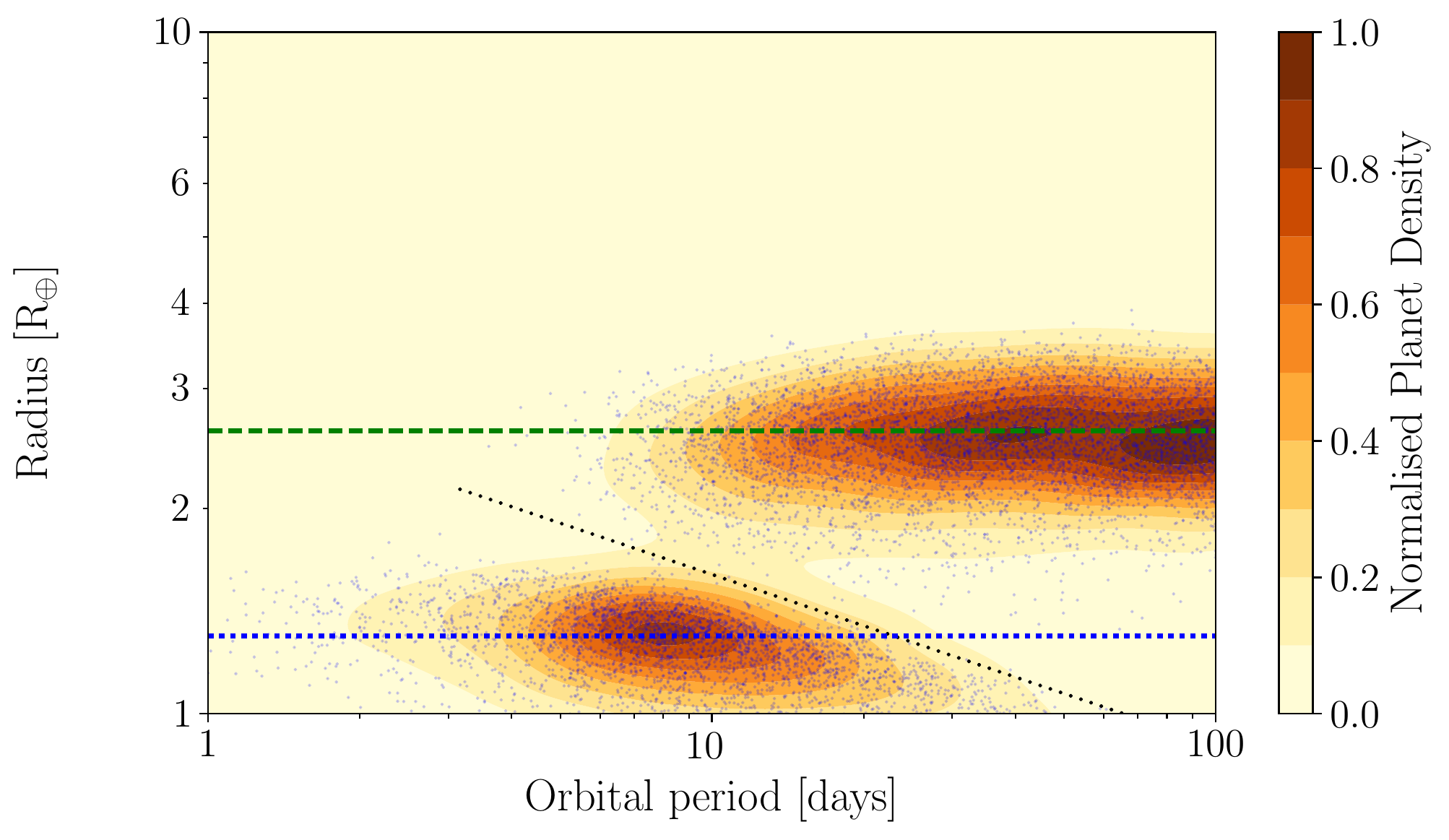}}\hskip0.4in {\includegraphics[width=8.5cm,trim=20 0 0 0,clip=]{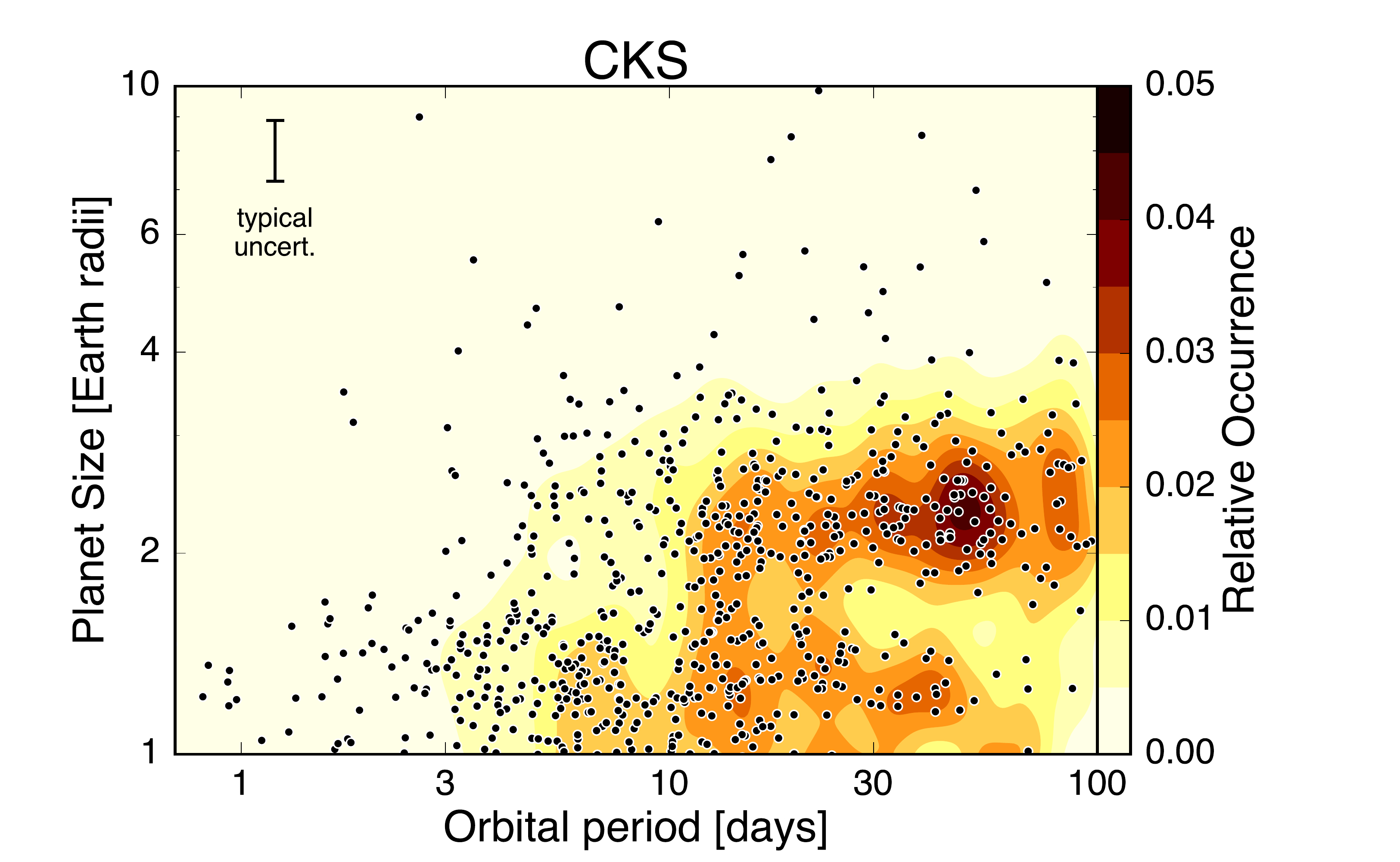}} \caption{The final radius distribution, now plotted as 2-D contours to display the period dependence. The model largely reproduces the observed one \citep[right side, taken from][with permission]{Fulton2017}, with the exception of an absence of small planets at long periods in the model.  The black dotted line is the analytical result (eq. \ref{eq:valley_comp}) for the size of the most massive planet that can be stripped bare, at a given period. This marks the lower 
boundary of the ``evaporation valley''.}  \label{fig:xvalley_2d} \end{figure*}

We evolve our initial population for three billion years under the influence of cooling and evaporation. The resulting radius distribution for planets with orbital periods less than 100 days is presented in Fig. \ref{fig:xvalley_rayleigh}. Compared to the observed 1-D radius distribution from the CKS sample, our model reproduces the observed features: the positions of the radius gap and the radius peaks, the widths of the peaks.  More significantly, the observed valley in the radius-period 2-D plane is reproduced nicely by our model. This is shown in Fig. \ref{fig:xvalley_2d}, alongside with the observed results by \citet{Fulton2017}. Both these successes strongly support the photoevaporation theory for the evolution of {\it Kepler} planets.

%Therefore, it is clear the ``evaporation valley'' first predicted in numerical models by \citet{Owen2013} %and now observed in the exoplanet population \citep{Fulton2017} is naturally explained by evaporation %driven evolution of low-mass exoplanets.

Since we have not attempted to ``fit'' the observed distribution, but rather to match its generic features, the agreements are not perfect. Short of theoretical predictions for the initial planet properties (how they depend on, e.g., separation, stellar mass, planet core mass, etc.), and only using an approximate photoevaporation model (the energy-limited case), this not surprising. In future work, using numerical models, one may actually be able to use the observed planet population to infer these initial properties. In the following, we discuss what parameters change may impact on the agreements.

%Small changes to our initial distributions (such as the initial envelope properties depending on orbital %separation and core mass, something expected theoretically e.g. \citealt{Lee2015}) and inclusions of %the missing physics described throughout the construction of our simple model does allow for a better %``fit''; however, the point of this work is to allude to the basic physics that controls the evaporation %valley and the observed bimodal distribution.

\section{Discussion}
We have used an analytic model to demonstrate that the evaporation valley is a robust outcome of the evolution of close-in, low mass planets with volatile envelopes. The mass-loss timescale always peaks for a H/He envelope of a few percent in mass, where the atmosphere roughly doubles the planet's radius \citep[see also][]{Chen2016}.
% This result is insensitive to both core mass and separation. Therefore,
This simple combination of planetary structure and evaporation is the origin of the bimodal radius distribution reported in \citet{Fulton2017}. Here, we explain these results in more depth, and investigate impacts on our model from various model uncertainties.

\subsection{The origin of the evaporation valley}

\begin{figure*}
\centering
\includegraphics[width=\textwidth]{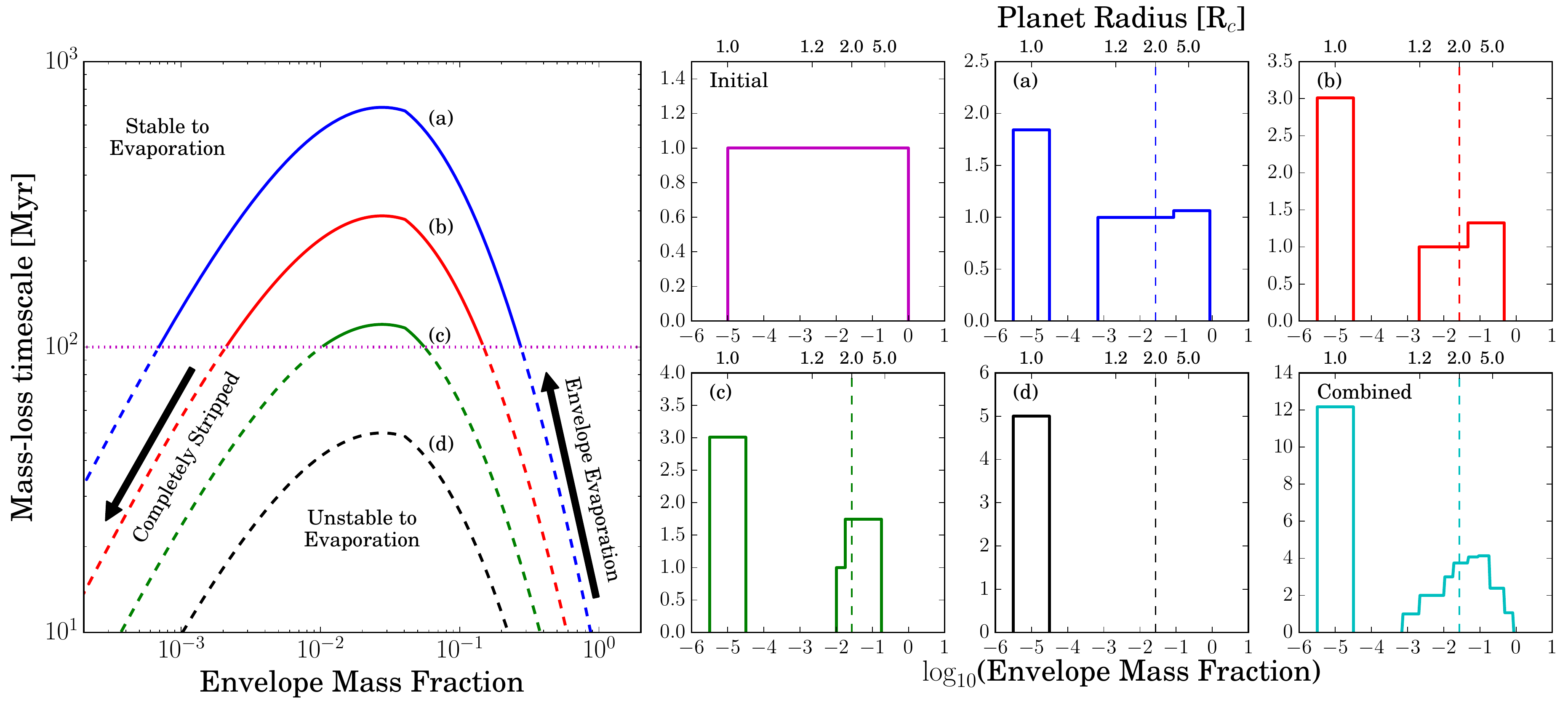}
\caption{Schematic figures showing the mass-loss timescale (far left panel) and resultant envelope mass fraction histograms that result from envelope evaporation. The far left panel show the mass-loss timescale as a function of envelope mass fraction for four models -- (a) through (d) -- which are progressively closer and closer to their parent star. Those envelope mass fractions with mass-loss times $<100$~Myr and unstable to evaporation and shown as dashed lines, whereas envelope mass fractions with mass-loss times $>100$ Myr are stable to evaporation and are shown as solid lines. %Planets whoes mass-loss time increases with decreasing envelope mass fraction can be unstable to mass-loss are will evolve towards the envelope mass fraction at which the mass-loss timescale is 100~Myr. Planets whoes mass-loss timescale decreases with decreasing envelope mass-fraction can undergo run-away mass-loss and be completely stripped leaving bare cores.  
The six small panels schematically shows what would happen to a population of planets. The top left small panel shows the initial envelope mass fraction distribution (arbitrary chosen to range between $10^{-5}$ and $1$). The panels labelled (a) through (d) shows the resultant population due to evaporation. %Planet's that the far left panel indicate are unstable to evaporation are either evaporated to the point where they are stable, or completely stripped (indicated by an envelope mass fraction of $10^{-5}$). Planet's that the far left panel indicates are stable are left at their initial envelope mass fractions. 
The bottom right panel shows the combination of models (a) through (d). The vertical dashed lines shows the envelope mass fraction which doubles the planet's radius. We clearly see how evaporation generates a bimodal distribution in radius and envelope mass fraction. 
%Planet's are either stripped cores or those with envelope mass-fractions that roughly double the planet's radius. {\w THIS DOESN''T JIBE with Fig. 4 where some initial x-values have to be excluded to make a gap.}
}\label{fig:diagram}
\end{figure*}

%We can understand the origin of the bifurcation in evolution shown in Figure~\ref{fig:xevolve} and the %resulting bimodal radius distribution by invoking our simple model. 
The origin for the evaporation valley is schematically shown in Fig.~\ref{fig:diagram}.  Consider a group of identical, low-mass planets differing only in their initial envelope mass fractions ($X_0$) { and high-energy exposure.} Since after the first 100 Myrs, both the stellar flux decays and the planet cools down to a smaller size, the evaporation is dominated by that in the early stages.  For a group of planets with the same high-energy exposure, if the peak mass-loss timescale  -- $t({\dot X})$ for $X=X_2$ -- is shorter than 100 Myrs (model d in Fig.~\ref{fig:diagram}), all envelopes are stripped bare and we expect to see only naked cores, as is the case for the observed Kepler planets at small separations \citep[e.g.][]{Dressing2015}. If, on the other hand, the peak mass-loss timescale is longer than $100$ Myrs (group c in Fig.~\ref{fig:diagram}), there is a bifurcation of final envelope masses -- planets with initial envelope masses $X<X_2$ will be completely stripped and present as naked cores; while planets with initial $X > X_2$ will be herded towards $X = X_2$. This manifests as a bifurcation in planet radius, with peaks at both the core size and its double. This bifurcation is the origin of the evaporation valley.

Lastly, for planets in group a and b which experience too little evaporation, there are little modifications to their atmospheres unless they start with extreme envelope masses (very high or very low). There is still a shepherding toward the above mentioned two peaks, but the widths of the second peak (at twice the core radius) are broader. The observed valley shape can then be used to exclude some of these models, as is done in Fig. \ref{fig:xgap}, where we show that most of the initial envelopes should have more than a few percent in mass. 

This understanding allows us to derive some useful scalings. These are detailed here and below.

Since less massive planets can be stripped out to larger distances, the bare population should have a decreasing size (smaller mass and hence smaller core radius) going away from the star. We can derive this by requiring that, for a given core mass, the longest evaporation time (when $X=X_2$, or $\Delta R \sim R_c$) is of order $t_{\rm sat}$ or 100~Myr,
\begin{equation}
t_{\dot{X}}(X=X_2,t=t_{\rm sat})\sim {t_{\rm sat}}\, ,
\label{eqn:cond1}
\end{equation}    
%Combining this with eqs. \refnew{eq:rhocore} \& \refnew{eq:tdotx}, we find that the most massive planet that can be stripped at a given period has a radius of
%\begin{equation}
%R_c \propto \eta^{0.18} M_*^{0.08} \rho_{M_\oplus}^{-0.03} P^{-0.25}\, .
%\label{eq:Rbare}
%\end{equation}
%This is drawn as a black dashed line in Fig. \ref{fig:xvalley_2d}. 
%{\w THIS DIFFERS FROM JAMES' RESULTS, KEPT HERE}
%Using eq. \refnew{eqn:cond1},
or, 
\begin{equation}
\frac{GM_p^2X_2}{8\pi R_c^3}\sim\eta t_{\rm sat} {{L_{\rm HE}}\over{a^2}} \approx\eta\mathcal{X}_{\rm HE}
\end{equation}
where $\mathcal{X}_{\rm HE}$ is the high-energy ``exposure'' for a given planet. Then combining eq. \refnew{eq:rhocore} \& \refnew{eq:X2value}, we can find the radius {\bc of} the most massive planet that can be stripped {\bc ($R^{\rm bot}_{\rm valley}$)} at a given exposure as:
\begin{equation}
R^{\rm bot}_{\rm valley}\propto \eta^{0.18}\,\mathcal{X}_{\rm HE}^{0.19}\,\rho_{M_\oplus}^{-0.24}\, . \label{eq:valley_comp}
\end{equation}
Using the high-energy flux dependence on stellar mass (\S \ref{sec:star_parameters}), ${\mathcal{X}}_{\rm HE} \propto M_*/a^2$, as well as eq. \refnew{eqn:X_R}, we convert the above relation to find $R^{\rm bot}_{\rm valley}\propto P^{-0.25}$, shown as the dotted black line in the left hand panel of Fig.~\ref{fig:xvalley_2d}. This explains the topology of the evaporation valley in the radius-period plane, in models with a constant evaporation efficiency $\eta$.  As we discuss later, this scaling can change when the evaporation model is different, while the core composition can shift the black line vertically.
%So an observational determination of this slope could be used to discriminate between evaporation %models. We also note that the core composition also plays an important role in setting the valley %position as we shall discuss in the next section. 

\subsection{Core Composition}\label{sec:core_composition}

When the evaporation valley was first predicted by \citet{Owen2013,Lopez2013} and {\bc subsequently by \citet{Jin2014}}, its location was conjectured to be a discriminant of core composition {which could provide clues as to their formation \citep{LopezRice2016}}. And indeed our model shows clear dependency on the core composition. 

\begin{figure} \centering {\includegraphics[width=0.4\textwidth,trim=95 190 50 120,clip=]{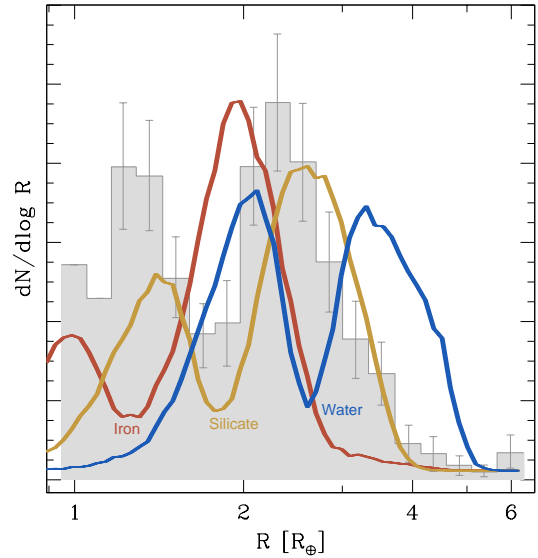}} \caption{Comparing the valley location for cores made up of pure iron ($\rho_{M\oplus}=11 {\rm ~g~cm^{-3}}$), pure silicate ($4 {\rm~g~cm^{-3}}$) and pure water ($1.3 {\rm ~g~cm^{-3}}$). All parameters are otherwise identical, including the core mass distribution. The left-ward shifting of the valley with rising density is as prescribed in eq. \refnew{eq:Rvalley}.  The observed data (grey shaded histogram) {\w excludes ice-rich cores (blue curves) and favours compositions that are roughly terrestrial, namely, silicate-iron composite.} }\label{fig:xvalley_composition} \end{figure}

Following our discussion above, we posit that the valley, independent of
the core composition, always lies in-between $R_c$ and $2 R_c$. Let us say, $\sqrt{2} R_c$. Or
\begin{equation}
R_{\rm valley} \sim 1.85 R_\oplus \left({{\rho_{M_\oplus}}\over{5.5 {\rm ~g~cm^{-3}}}}\right)^{-1/3} \left({{M_c}\over{3 ~{\rm M}_\oplus}}\right)^{1/4} \, .
\label{eq:Rvalley}
\end{equation}
This is demonstrated in Fig.  \ref{fig:xvalley_composition}, where cores of the same mass but different composition exhibit different valleys. An order of magnitude in change in $\rho_{M_\oplus}$
% due to a change in core composition from entirely Iron to entirely Ice/water moves the %position of the evaporation valley by a factor of $\sim 2$ in radius, 
leads to an easily detectable shift in the valley position of a factor 2, much larger than the $10\%$ radius errors achieved in the CKS sample \citep{Johnson2017}, and even large compared to the $30\%$ error in the general {\it Kepler} sample.  { So if the planet mean masses are known, the valley position yields the core composition.} 

{Interestingly, even when the planet mean masses are not known, if we assume the bimodal radius distribution arises from a single population of planets, we can break the degeneracy and determine the composition.  In our model, we successfully reproduce the observed valley using a single population of planets with $M_c \sim 3 ~{\rm M}_\oplus$ (see also \S\ref{sec:two_pop}) and terrestrial compositions. This does not seem possible for {\w some} other compositions. For instance, if the cores are made up largely of ice/water with $\rho_{M\oplus} = 1.3\, {\rm g~ cm^{-3}}$, the bare cores at $R=1.3~ {\rm R}_\oplus$ will correspond to $M_c \sim 0.5 ~{\rm M}_\oplus$, and none of these cores can retain enough hydrogen to occupy the second radius peak. Similarly, for pure iron composition ($\rho_{M\oplus}=11\, {\rm g~ cm^{-3}}$), the bare cores should correspond to $M_c \sim 6~ {\rm M}_\oplus$.  However, few of these planets could be evaporated down to naked cores, at the distances that we observe them. } So the current data {\w exclude cores that are mostly icy, and favour compositions that are terrestrial-like, i.e., silicate-iron composite. This is} consistent with RV results from known bare planets \citep[e.g.][]{Dressing2015}.

\begin{figure} \centering {\includegraphics[width=0.4\textwidth,trim=90 150 50 148,clip=]{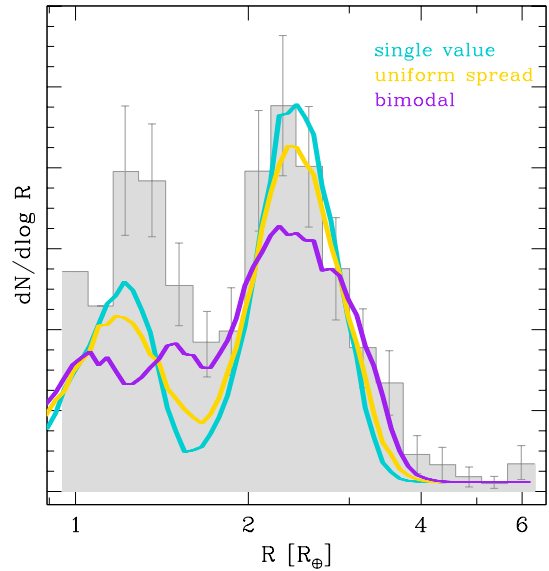}} \caption{Same as Fig. \ref{fig:xvalley_composition}, but now focussing on silicate-iron composites with different iron fractions ($f_{Fe}$). Different theoretical distributions correspond to models with: single value, $f_{Fe} = 1/2$; uniform spread, $f_{Fe} \in [0,1]$; bimodal, $f_{Fe} =0$ or $1$. The data excludes the last distribution, but can not distinguish the first two. This illustrates our inability to constrain the iron fraction to a narrow range. }\label{fig:xvalley_ironfraction} \end{figure}

{\w While these cores are Earth-like, we find that we cannot constrain their iron fractions to a narrow range using current data: in Fig. \ref{fig:xvalley_ironfraction}, we present the expected size distributions for different iron fractions, and all appear largely consistent with data.
  This could reflect either a genuine composition spread in real planets, or an intrinsically narrow composition spread that is smeared out by errors in the observed data.}

% Lastly, we can say something about the homogeneity of the core composition. If our population of planets are made up of planets with a range of compositions, say, with $\rho_{M_\oplus}$ varying by a factor of $2$,\footnote{For example, for a silicate-iron composite body, this corresponds to varying the iron fraction from $0$ to $70\%$ \citep{Fortney2007}.} the valley positions for each core composition would have shifted by more than $20\%$ within the population, erasing the observed valley between $1.6$ to $2 ~{\rm R}_\oplus$.

In conclusion, the observed gap {\w suggests that the most common core composition is Earth-like. But the iron fractions in these silicate-iron composite could span a wide range; with a detailed numerical fit to the observed population one maybe able to constrain the Iron fraction to a narrower range.}

\subsection{Core Luminosity}
\label{sec:core_lum}
For simplicity, in the main  analysis, we have ignored the luminosity contribution from the cooling core. This is not justified in general, since the thermal content in the more massive core can be more substantial than that in the envelope. We address this deficiency here.

We assume that the entire planet, core and atmosphere, cools with a single temperature. This is reasonably accurate for the solid interior, since even the adiabatic temperature gradient is relatively shallow.  For an envelope with $\Delta R \lesssim 4 R_c$, ($X\lesssim 0.1$) the thermal inertia is dominated by gas in the bottom few scale heights close to the core. Gas here likely share the same temperature as the solid part.  We further assume that the core contributes only through its primordial heat, and ignore any radiogenic source. This is valid, especially at early times.  For the heat content of the core, we adopt a heat capacity of $10^7$ erg g$^{-1}$ K$^{-1}$ \citep[][similar to rock]{Valencia2010},
and we assume that the core is made up of SiO$_2$ molecules with a
mean molecular weight of $76$ (or per particle specific heat of $\sim 7 k_B$). The envelope is assumed to be  composed of H-atoms with a per particle specific heat of $3/2 k_B$.  
So including the core contribution to the internal luminosity, we have
\begin{eqnarray}
L & = & L_{\rm core} + L_{\rm env} \approx \left[1 + {{7k_B \times ({M_c}/{76 m_H}})\over{1.5 k_B \times ({XM_c/m_H})}} \right] L_{\rm env}\nonumber \\
&  \approx & \left(1 + {1\over{17 X}}\right) L_{\rm env}\, .
\label{eq:coreheat}
\end{eqnarray}
This expression explains the results shown in Fig.~3 of \citet{Lopez2014}.  We insert this factor into eq. \refnew{eq:eqL} and re-calculate the photo-evaporation process. Core luminosity allows the adiabatic envelope to remain large for a longer time, enhancing mass-loss. However, the overall effect is minor. This is because the evaporation bottleneck is when the envelope mass fraction is of order a few percent ($X = X_2$). At this point, the core contribution to the luminosity is  an order unity effect.

\subsection{Atmosphere Metallicity}
\label{subsec:zdep}

Some studies of transmission spectroscopy have suggested that the atmospheres of hot Neptunes (e.g., GJ1214b) are highly enriched in metals, with [Z/H] perhaps as high as 100 \citep[e.g.][]{Charnay}, {while others (e.g. HAT-P-26b) are consistent with solar metallicity \citep{Wakeford2017}} .  Here we explore how atmosphere metallicity may affect our results.

Metallicity enters our model in two ways (eq.~\ref{eqn:X_R}). First, higher metallicity increases the mean molecular weight ($\mu$), which reduces the scale height at a given temperature. This shrinks the atmosphere, and increases the value of $X_2$. Second, higher metallicity increases opacity ($\kappa_0$), which shifts the convective-radiative boundary. Thus, the atmosphere looks more inflated for a given age. The outcome of our model is determined by the competition between these two effects \citep[also see][]{Howe2015}.
% their fig. 11
Reality may be more complicated. The constant energy-efficiency approach we adopt here may be severely invalid if metals affect the driving of the photo-evaporative flow.\footnote{For X-ray driven flow, metals (like C, O) dominate both the absorption of X-ray photons and the line cooling of the heated region \citep{Owen2012}. The two effects may cancel out to a large degree. For EUV driven flow, hydrogen photo-ionization and recombination dominate the heating and cooling, though metals, if present at a large quantity, may reduce the energy efficiency  as additional sources of radiative cooling.}  Metals may have additional thermal effects by, for instance, producing high altitude hazes/clouds.

%Defining metallicity $Z$ to be the mass ratio of metals relative to hydrogen, with the metallicity of the Sun %being $2\%$, we find that raising $Z$ by a factor of $100$ from the Solar value raises the mean molecular %weight (of a molecular mix) from $2.35$ to $5.52$. In the meantime, $\kappa_0$ is roughly increased by %a factor of $33$. So the two cancels.
Assuming that $\kappa_0 \propto Z$ \citep[e.g.][]{Lee2015}, we find that, overall, raising the metallicity leads to a slightly larger radius and hence a slightly shorter mass-loss timescale. The effect maximizes  (making a factor of 2 difference in timescale) when the metallicity is $10\times$ solar, but nearly vanishes when it is $100\times$ solar. Such an effect is so small it is eclipsed by other uncertainties in our current model. As a result, we can not make conclusive statement regarding atmosphere metallicity, aside from the fact that the initial atmosphere must be dominated by H/He in number (and so in molecular weight). For example, the observed position in the radius-period plane rules out the presence of ``water-worlds'' whose volatile envelopes are primarily water/steam. Water/steam atmospheres have a much larger mean-molecular weight ($\mu\sim 18$) compared to H/He envelopes, this significantly increases the value of $X_2$ to values more like $\sim 0.5$ \citep{Lopez2016}. The increased high-energy exposure required to evaporate an envelope with $X=0.5$ (even with an optimistic efficiency of 0.1 for steam atmospheres) results in the evaporation valley appearing at much shorter separations, around a period of 2~days \citep{Lopez2016}, rather than the 10~days it is observed. Therefore, water-worlds making up any reasonable fraction of the {\it Kepler} planets is clearly ruled out by the CKS observations.    

%\subsection{Using the evaporation valley to test evaporation models}
\subsection{Evaporation Efficiency}
\label{subsec:model2}\label{sec:evap_model}

There are two reasons why the evaporation efficiency may not be constant for all planets. First, geometry. We define the efficiency $\eta$ to be based on the light received by the planet disk ($R_p$). However, the true cross-section is determined the {UV/}X-ray photosphere. This lies well above the planet photosphere, especially so for planets with large surface scale heights. This raises the efficiency for those planets.  Second, physics.
% \x{As indicated by Equation~\ref{eq:valley_comp}, not only is the position of the evaporation valley sensitive to core composition, it also predicts that the planet radius at which the occurrence valley appears should increase with increasing X-ray exposure. As one moves to smaller and smaller separations evaporation can completely strip even more massive (and larger) cores. As pointed out by \citet{LopezRice2016} this slope is a unique prediction of evaporation driven evolution. However, we note that while the valley's radius should increase with increasing x-ray exposure the steepness of the slope is sensitive to the evaporation model. The scaling given in Equation~\ref{eq:valley_comp} is only valid for a constant efficiency ``energy-limited'' model. In fact}
Most hydrodynamic models predict that the efficiency should drop as one moves to more massive and/or denser planets. The deeper gravitational potential in this case means it takes the flow a longer time to escape, allowing it to lose more energy by radiative cooling \citep[e.g.][]{Owen2012}.  The ``energy-limited'' evaporation is only applicable to UV evaporation of weakly irradiated planets (where recombination-equilibrium cannot be reached, and radiative cooling is inefficient on flow-timescales) with low escape velocities \citep{OA16}, but not the full range of planets observed to make up the evaporation valley, where recombinations \citep[e.g.][]{MurrayClay2009} and X-ray evaporation \citep[e.g.][]{Owen2012} complicate the picture. %\x{For example, an efficiency that decreases as the core mass (and core radius) increases results in a much flatter evaporation valley than the energy-limited model (Equation~\ref{eq:valley_comp}) predicts.}

We use a simple model to illustrate these effects. We assume that the efficiency of mass-loss scales as\footnote{We caution the below expression is only suitable in a narrow range of parameter space. We use it here for illustrative purposes and it should {\it not} be used to describe the results from radiation-hydrodynamic evaporation models.} 
\begin{equation}
\eta = 0.1 \left(\frac{v_{\rm esc}}{15 {\rm km\,s^{-1}}}\right)^{-2} \, ,
\label{eq:etanew}
\end{equation}
where $v_{\rm esc}$ is measured at the planet photosphere and the normalization of $15{\rm ~km~s^{-1}}$ is chosen to reproduce the observed distribution around 10 days. The power $2$ roughly coincides with the result shown in Fig. 13 of \citet{Owen2012}, where $\eta \propto M_c^{-3/4}$ for earth-like cores in the range 1-10\,M$_\oplus$. {\bcc We adopt this efficiency scaling over other published scalings \citep[e.g.][]{Salz2016} as it is more appropriate for young, active stars which drive most of the mass-loss, rather than older stars where evaporation does not affect a planet's evolution.}

\begin{figure} 
\centering
% run planet_ensemble.f with eta_treatment set to .true.
%{\includegraphics[width=0.49\textwidth,trim=120 760 65 200,clip=]{x_etanew.png}}
{\includegraphics[width=0.49\textwidth,trim=0 0 0 0,clip=]{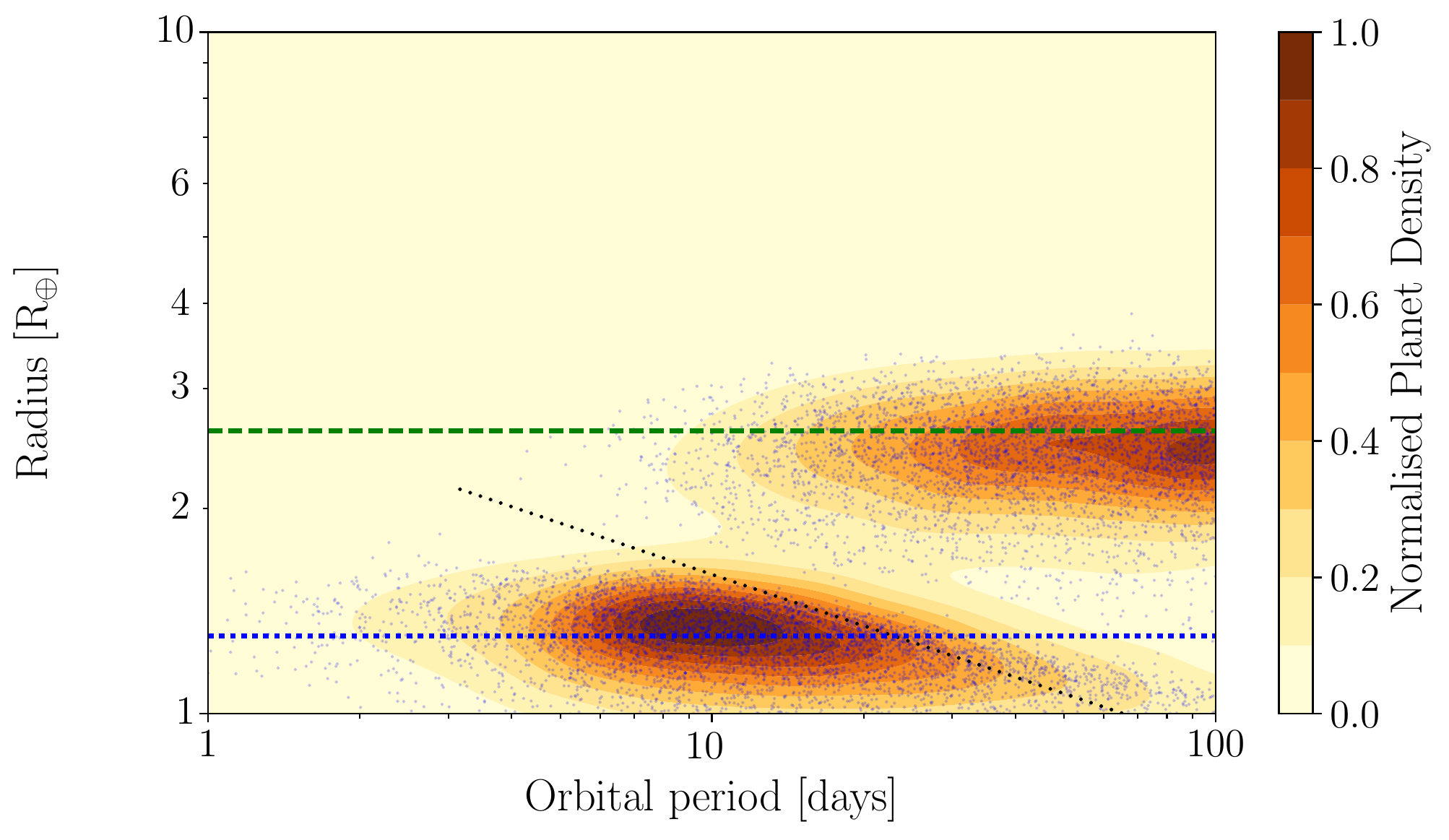}}
\caption{Similar to Fig. \ref{fig:xvalley_2d} but with an evaporation efficiency that is larger for less bound envelopes. (eq. \ref{eq:etanew}). In this case, light planets can be stripped out to larger orbital distances, producing an extended tail of small planets out to $\sim 100$ days. For comparison, the dotted black line is eq. \refnew{eq:valley_comp}, applicable for models with a constant $\eta=0.1$.}
\label{fig:etanew}
\end{figure}

With such an ad-hoc efficiency law, low-mass planets can be stripped out to larger distances, while high-mass planets are more resistent to stripping. Re-working through the derivation leading to eq. \refnew{eq:valley_comp}, we obtain $R^{\rm bot}_{\rm valley} \propto P^{-0.16}$, a shallower dependence than
that in the constant efficiency model. This is illustrated in Fig. \ref{fig:etanew}, where one observes 
an extended tail of small-planets towards long periods. 

%{\y The lower efficiency for more massive planets also means that they can only be stripped at closer %distances. This results in an evaporation valley that is flatter than the one expected from the %energy-limited model. Repeating the analysis in Section~\ref{sec:core_composition} but including %Equation~\ref{eq:etanew} results in a radius-period scaling for the bottom of the evaporation-valley of %$R_{\rm val}^{\rm bot}\propto \mathcal{X}_{\rm HE}^{-0.12}\propto P^{-0.16}$. Combining %Equations~\ref{eq:valley_comp} and equation \ref{eq:etanew} we can demonstrate the combined effects %of changes in efficiency and core composition on the position of the evaporation valley. 

\begin{figure}
\centering
\includegraphics[width=\columnwidth]{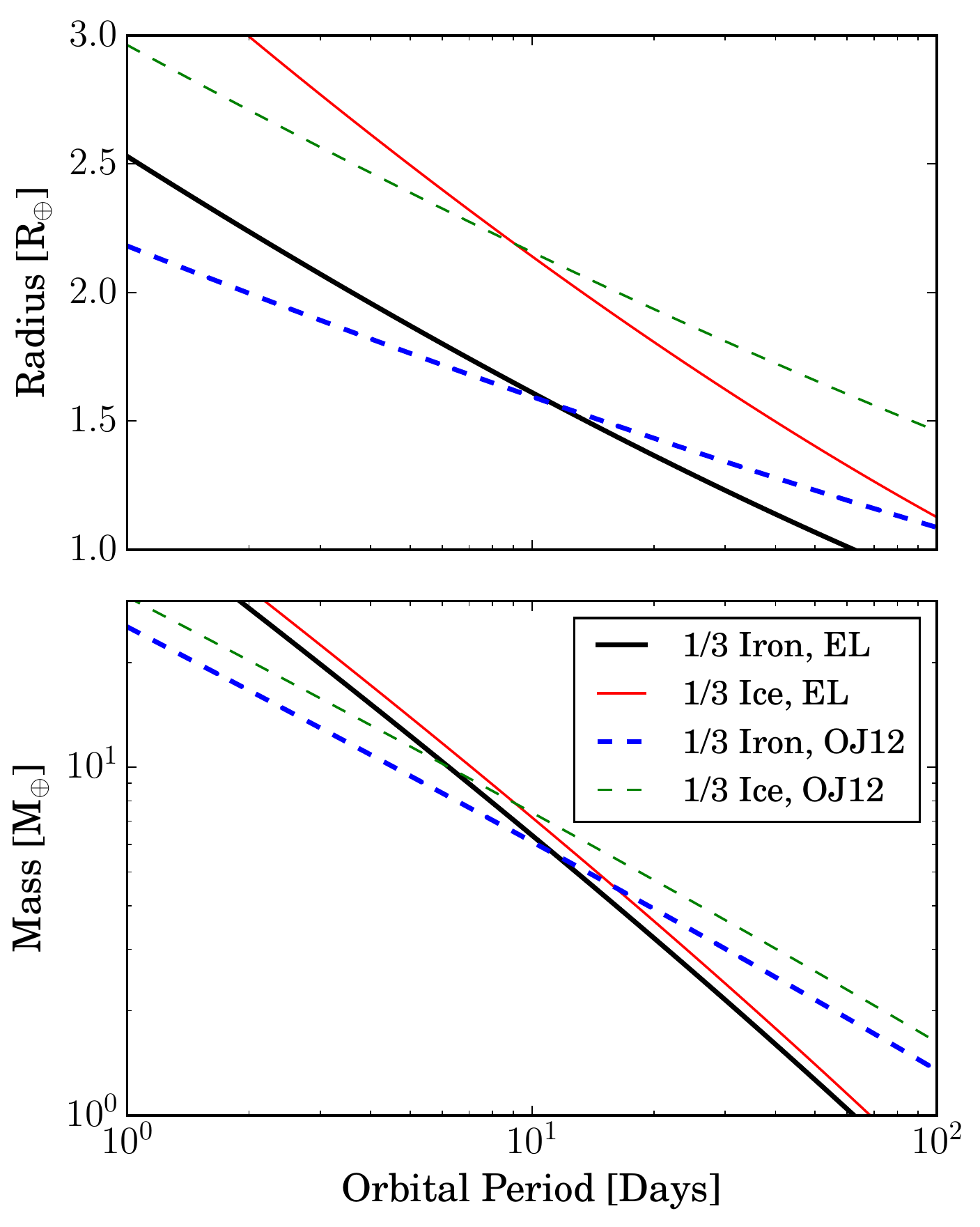}
\caption{The position of the bottom of the evaporation valley (i.e. the largest stripped cores at a given period) shown in the top panel, with the core masses shown in the bottom panel. Solid lines show constant efficiency energy-limited models while dashed lines show evaporation models with variable efficiency similar to the \citet{Owen2012} evaporation models. The thick lines show Earth-like composition cores, while the thin lines shown those composed of 1/3 ice and 2/3 silicates.}\label{fig:valley_pos}
\end{figure}

To further illustrate this effect, in Fig. \ref{fig:valley_pos}, we plot the values of $R^{\rm bot}_{\rm valley}$ for Earth-like cores and those with a composition of 1/3 Ice, 2/3 silicates, using the \citet{Fortney2007} mass-radius relationship. This figure demonstrates that, with a large enough sample of planets, and a $\sim 10$\% radius errors,  one could in principle distinguish between different evaporation models and different core compositions. The bottom panel translates the radius into planet core mass, demonstrating how the mass of the most massive stripped core depends on orbital period, for any of the models. Follow-up mass measurements can test these predictions.
%However, an Earth (1AU around the Sun) is unlikely to get rid of a percent-level hydrogen envelope %\citep{REFS??}.

%resulting in a valley that scales with X-ray exposure as $R_{\rm val}\propto\mathcal{X}_{\rm HE}^{0.12}$. 

{\w Finally, we have not considered non-thermal escape processes such as stellar wind stripping. This is because hydrodynamic evaporation dominates for the separations we are interested in here \citep[e.g.][]{MurrayClay2009}. We have also not included the effects of a planetary magnetic field. Presence of a magnetic field strong enough to retain its dipolar structure (despite the evaporation) close to the planet may suppress the mass-loss rates significantly \citep[e.g.][]{Adams2011,OA14,Khodachenko2015,OAd16}. Unfortunately, it is unknown if the {\it Kepler} planets possess sufficiently strong fields.}

\subsection{Stellar Mass}

The deficit of planets between $R_c$ and $2 R_c$ results from interaction between planet internal structure and photo-evaporation. It is not sensitive to stellar mass. Indeed \citet{Fulton2017} report that the valley remains the same for all stars in their sample.

On the other hand, the actual morphology of the valley, i.e., the relative height of the two peaks, the widths of the peaks, the shape of the valley in period space, etc., depends on stellar mass.
When plotting planet radii for a population that have a spread in host masses, it is easiest to see the evaporation valley when the horizontal axis is chosen to be the period, {not bolometric insolation as commonly assumed}. As eq. \refnew{eq:tdotx} shows, if we assume that
$L_{\rm sat} \propto M_*$, the evaporation timescale scales with orbital period and stellar mass as,
\begin{equation}
t_{\dot X} \propto P^{1.41} M_*^{-0.48}\, ,
\label{eq:tdotx2}
\end{equation}
i.e., a weak dependence on stellar mass. Alternatively, if one takes the horizontal axis as the stellar insolation ($I \propto L_*/a^2$, or equivalently, planet equilibrium temperature),
\begin{equation}
t_{\dot X} \propto I^{1.06} M_*^{2.2}\, .
\label{eq:tdotx3}
\end{equation}
This latter has a stronger mass dependence; as a result, the evaporation valley is less distinct \citep[Fig. 10 of][]{Fulton2017}. Similarly, but to a lesser degree, is plotting by semi-major axis, with $t_{\dot X} \propto a^{2.12}M_*^{-1.19}$. Ultimately, the stellar mass independent variable to use is high energy exposure (the total received high-energy flux over the planet's lifetime); however, the down side is high energy exposure is a model dependent quantity and not observationally accessible for individual planets. The current high energy flux received by a planet at billions of years evolution is not representative of {\bc its} high energy exposure \citep[e.g.][]{Tu2015}.

% see x_mstar.png
%Here, we are most concerned about sun-like stars. How does the evaporation valley look for M stars, the 
%main-stay of the {\it TESS} mission? The position of the evaporation valley is likely independent of the %stellar mass, as we argue here. These stars have longer activity lifetimes \citep{West}, of order 1 Gyrs to a few Gyrs. 

\subsection{Two populations of planets?}
\label{sec:two_pop}

Envelope stripping becomes decreasingly important as one moves away from the star.
For our assumed population of gas-rich planets, this means, going outward, there should be fewer bare cores, and these bare cores should be lower in mass (smaller in size). 
{\bc For example,  Fig. \ref{fig:xvalley_2d} (the dotted black line) {\bc indicates} that there should be no bare planets of mass $3~ {\rm M}_\oplus$ ($R_c = 1.3 ~{\rm R}_\oplus$) beyond orbital periods of $30$ days and no bare planets of mass $1\,$M$_\oplus$ beyond orbital periods of $60$ days} {\citep[see also][]{LopezRice2016}}.\footnote{The variation of $\eta$ (\S \ref{subsec:model2}) may extend this value out to $50$ days (Fig. \ref{fig:etanew}) for bare 3\,M$_\oplus$ planets and out to $100$ days for  bare 1\,M$_\oplus$ planets.}  Furthermore, \citet{Owen2012} showed that the hydrodynamical outflow we posit here do not happen for planets much outside $30$ days. Rather, 
%\citet{Owen12} mass-loss suggests this typically happens around 0.2-0.3~AU and %would correspond to a radius range of around $\sim1.2$~R$_\oplus$ for Earth-like %composition cores.
the evaporative outflows is so rarefied the gas stops being collisional before the sonic point is reached and 
%\footnote{Since anything that happens beyond the sonic point is no longer in casual contact with the planetary atmosphere it does not matter if the gas is no longer collision beyond the sonic point.} 
the mass-loss rate is instead determined by Jeans' escape and falls much below the hydrodynamic value. 
%We do not expect to see $3 M_\oplus$ stripped planets much beyond $30$ days, or $1 %M_\oplus$ planets beyond $\sim 100$ days. 

{\w Interestingly}, despite the selection effects against detecting small planets at large distances, they appear to be present in the {\it Kepler} sample.\footnote{The CKS sample does not contain enough detection in this range, and it is unclear what the stellar masses of individual planets are.}  The radius error bars remain large. And the abundance of such planets are not yet solidly established. However, if future observations confirm that such planets are abundant, this will require a separate population of planets than that posited here. This population is born bare.  %Therefore, searches for truly terrestrial planets, ones that formed like the solar-system terrestrial planets after the gas disc dispersal, should focus on small planets with $>50~$day periods.
%However, if one were to introduce two populations, guided by results in Fig. %\ref{fig:xgap}, one could obtain much better similarity to the observed pattern, at least %superficially. Fig. \ref{fig:xvalley_2pop} shows one such example. 

%\subsubsection{Disappearance of valley at large separations}
%\subsubsection{Beyond 100 days}
%\label{subsec:largea}

%\subsection{A window into planet formation}

%-rocky cores 

%-formed inside snowline

%-formed in gas disc

%-at their current orbital periods by 100 Myr (no high e migration)

\section{Conclusions}

We have developed a {\bc minimal} analytical model which allows us to efficiently follow the evolution of low-mass planets, under the combined effects of cooling contraction and mass loss by evaporation, aside from illuminating the controlling physics. This model show that the mass-loss timescale peaks at around where planet sizes are doubled by their H/He envelopes, also where the envelope mass is of order a few percent. The timescale drops below this value because, while the envelope becomes more tenuous, the planet radii remain largely constant and so do the photo-evaporating fluxes they receive. The timescale also drops above this value because the planet swells up faster than the addition of envelope mass.

As a result, photoevaporation naturally gives rise to a final planet distribution that is bimodal in radius, peaking at the naked core size and twice its value. This then explains the observed radius ``valley'' in the California Kepler Survey sample \citep{Fulton2017}, as well as the steep fall-off of planets beyond $\sim 3 R_\oplus$ in the general {\it Kepler} catalogue, for a single population of planets that are born with 
% 0) trivial one but worth stressing: the valley is a feature that uniquely confirms that photoevaporation is the mechanism, as opposed to, say, different formation environments, to explain the observed correlation between planet size and distance.
at least a few percent of H/He envelopes (``water-worlds'' with water/steam envelopes are ruled out), this amount of gas lies above the most optimistic estimate for outgassing \citep{ElkinsTanton2008a,ElkinsTanton2008b} and suggests that the envelopes were accreted from the protoplanetary discs. In the latter scenario, a few percent or more envelope mass may be natural \citep{Rafikov2006}. It is roughly the mass of an adiabatic envelope maintained by a planet of a few Earth masses at $0.1$ AU \citep[but see][for modifications]{LeeChiang,Ginzburg2016}.   
%The upper-limit in gas fraction may contain imprints of the early ``boil-off'' process %\citep{OW16}.

The positions of the peaks and valley also lead us to the following conclusions: the planet masses can be described by a Rayleigh distribution with a mode at $3 ~{\rm M}_\oplus$, the cores have a composition {\w that is ice-poor. Ice-rich cores move the gap too far above the observed value.} The absence of icy cores indicates local assembly of the {\it Kepler} planets, as opposed to formation beyond the ice-line (followed by large-scale migration). Furthermore, as most of the erosion occurs before 100 Myrs, the {\it Kepler} planets must have reached their current orbital locations well before that time. This rules out the possibility that they were migrated in at late times by dynamical processes (e.g. high-eccentricity migration) and also rules out the suggestion that the dearth of intermediate sized planets on short orbits (the ``sub-jovian pampas'') is entirely due to tidal stripping during dynamical migration \citep{Matsakos2016}. {\w Lastly, current data allow the silicate-iron cores to have a wide spread in iron fraction.}

Planets that can be stripped bare have vanishingly small mass at larger orbital periods. Our energy-limited model does not produce planets of size $\sim 1.3 ~{\rm R}_\oplus$ (or $3~ {\rm M}_\oplus$) much beyond $30$ days.  We investigate the impacts on this (and other) prediction by factors such as atmosphere metallicity, core luminosity, stellar mass, and evaporation efficiency. None made much difference except perhaps the last one. In our ad-hoc model where the efficiency is higher for less bound atmospheres, we are able to extend the above period limit to $\sim 50$ days. This gives hope that a large enough observational sample could discriminate various evaporation models. 

Observationally, there is some evidence small planets appear to exist at orbital periods longer than $50$ days in the overall {\it Kepler} sample. They are possibly numerous given their low detection probability. If this is confirmed by future observations, then our one-population model, though appearing to explain many of the observed features inward of $100$ days, may be insufficient. There may need to be another population of low-mass planets that are born with essentially no envelopes. This separate population will need to have masses $M \leq 3 ~{\rm M}_\oplus$, so as not to fill in the evaporation valley.  Orbital instabilities have been suggested to be common among {\it Kepler} systems \citep[e.g.,][]{PuWu}. {\bcc Giant impacts among short period planets  will happen at speeds well above their surface escape velocities \citep{AgnorAsphaug,Marcus2009}, and will therefore disperse much of the original planetary envelopes. Since these impacts will occur after the natal disk has dispersed, they may give rise to this new population.}
%he CKS %data provide, from a different perspective, a hint for such an event.
%giant impact too random, also no clear prediction?

Future improvements are needed to further solidify results here.  Theoretically, our results are obtained using a {\bc minimal} analytical model. This was a deliberate choice in order to provide a basic understanding for the origin of the evaporation valley. We have calibrated our model against {\it mesa} calculations, but it still contains a number of assumptions (e.g., constant evaporation efficiency). Work should be carried out with more accurate numerical planetary structure models and more physically motivated evaporation model to determine the exact nature of the birth properties of short-period, low-mass exoplanets. Observationally, planet properties will be refined further by future precision stellar data, allowing more detailed comparison.  We have adopted a mass function for the {\it Kepler} planets that is Rayleigh with a mode of $3 M_\oplus$. This {prediction} should be tested by future mass measurements. 

{\w Notes in proof. While this paper was in review, \citet{Jin2017} submitted a paper which used numerical modelling to reach much of the same conclusions about the composition of the {\it Kepler} planets as we do here. An interesting development comes from \citet{DongLamost} where they reported a population of ``Hop-tunes'', Neptune-sized planets at close distances from their host stars, where we predict most planets should have been evaporated to bare cores. Intriguingly, this population only exists around metal-rich stars. This presents currently an unsolved puzzle.}

\acknowledgments { The authors are grateful to the referee for comments which improved the manuscript. We thank Nikhil Mahajan for performing some of the early explorations. We are grateful to BJ Fulton for permission to reproduce his figure and useful discussions. We acknowledge Tim Morton, Lauren Weiss and Josh Winn for helpful insights. }
JEO acknowledges support by NASA through Hubble Fellowship grant HST-HF2-51346.001-A awarded by the Space Telescope Science Institute, which is operated by the Association of Universities for Research in Astronomy, Inc., for NASA, under contract NAS 5-26555. YW acknowledges support from NSERC.

\appendix
\section{The dimensionless integrals $I_1$ and $I_2$}
\begin{figure}[h] 
\includegraphics[width=\textwidth]{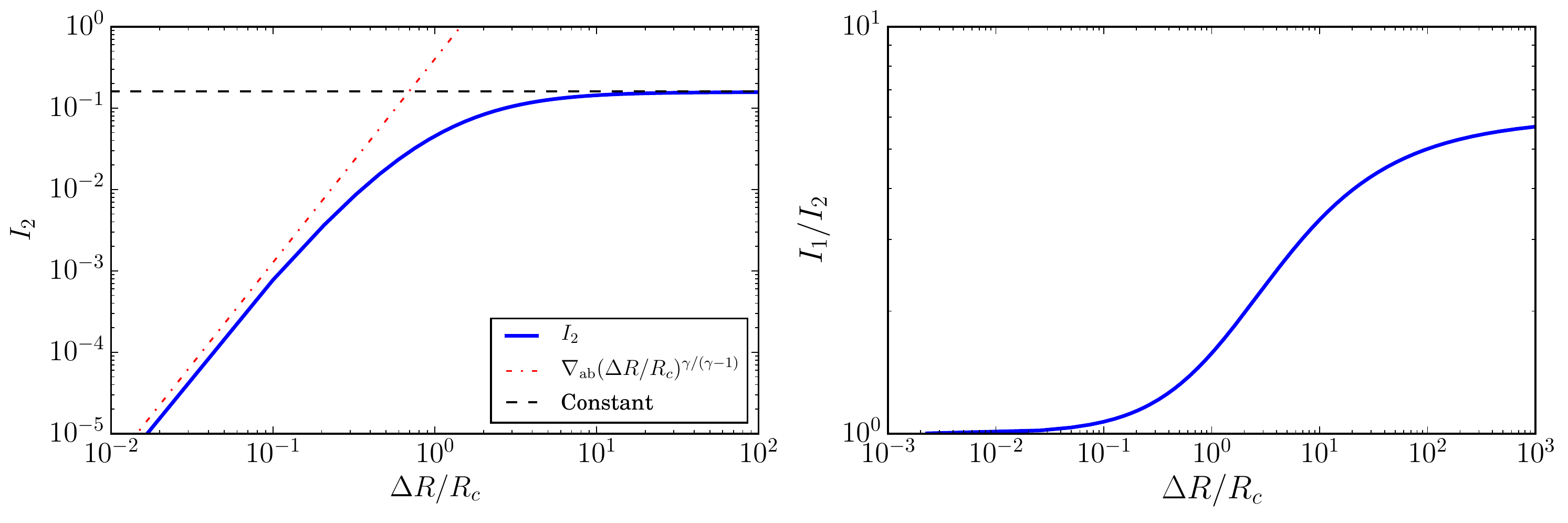}
\caption{The numerical solutions of $I_2$ (left-panel) and $I_1/I_2$ (right-panel) are shown as the solid blue line. The approximations to $I_2$ used in the analysis in \S\ref{sec:convective} in the limits $\Delta R <1$ (red dot-dashed) and $\Delta R > 1$ (black dashed) are shown in the left-hand panel. }\label{fig:integrals}
\end{figure}

In \S~\ref{sec:convective} we have introduced dimensionless integrals of the form:
\begin{equation}
I_n=\int_{R_c/R_p}^1x^n\left(x^{-1}-1\right)^{1/(\gamma-1)}{\rm d}x\label{eq:integral_dim}
\end{equation}
In the limit $\Delta R \ll 1$, or $x \sim 1$ then $I_n$ is independent of $n$ and can be approximated as:
\begin{equation}
I_n\approx\int_{R_c/R_p}^1\left(x^{-1}-1\right)^{1/(\gamma-1)}{\rm d}x\approx\nabla_{\rm ab}\left(\frac{\Delta R}{R_p}\right)^{\gamma/(\gamma-1)}
\end{equation}
This result implies that the ratio $I_1/I_2\approx1$ for the case of a thin envelope. For the case of a puffy envelope, we need to consider how the $I_n$ varies with $R_c/R_p$. For $\gamma=5/3$, the choice appropriate in our model, the integrand of eq.~\refnew{eq:integral_dim} is, to first order, $\propto x^{n-3/2}$ at small $x$. Therefore, for $n>1/2$ the integral is not dominated close to $x\sim R_c/R_p$ and eq.~\refnew{eq:integral_dim} becomes independent of $R_c/R_p$ as the envelope becomes large and $R_c/R_p$ becomes small. Thus in the limit of large atmospheres ($\Delta R>1$), both $I_1$ and $I_2$ are constant and as such their ratio $I_1/I_2$ is also constant. In our model we require $I_2$ and the ratio $I_1/I_2$ explicitly. The full numerical solution of $I_2$ and the approximations discussed above are shown in the left-hand panel of Figure~\ref{fig:integrals} and the full numerical solution of the ratio $I_1/I_2$ is shown in right-hand panel Figure~\ref{fig:integrals}. 

\end{document}